\documentclass{article}

\usepackage[top=3cm, bottom=3cm, left=3cm,right=3cm]{geometry}
\usepackage[colorinlistoftodos]{todonotes}
\usepackage{graphicx}
\usepackage{amssymb}
\usepackage{amsmath}
\usepackage{gensymb}
\usepackage{todonotes}
\usepackage{pdflscape}
\usepackage{caption}
\usepackage{subcaption}
\usepackage[sort,round]{natbib}
\usepackage[T1]{fontenc}
\usepackage[utf8]{inputenc}
\usepackage{authblk}
\usepackage{gensymb}
\usepackage{pdfpages}
\usepackage{setspace}
\usepackage{booktabs}
\usepackage{longtable}
\usepackage{float}
\usepackage{tikz}
\usepackage[colorlinks=true,citecolor=blue, linkcolor=blue]{hyperref}

\def\ci{\perp\!\!\!\perp}

\title{Data integration for high--resolution, continental--scale estimation of air pollution concentrations}

\author[1]{Matthew L. Thomas}
\author[2]{Gavin Shaddick}
\author[3]{Daniel Simpson}
\author[4,5]{Kees de Hoogh}
\author[6]{James V. Zidek}
\affil[1]{Department of Infectious Disease, Imperial College London, London, U.K.}
\affil[2]{Department of Mathematics, University of Exeter, Exeter, U.K.}
\affil[3]{Department of Statistical Sciences, University of Toronto, Toronto, Canada.}
\affil[4]{Swiss Tropical and Public Health Institute, Basel, Switzerland.}
\affil[5]{University of Basel, Basel, Switzerland.}
\affil[6]{Department of Statistics, University of British Columbia, Vancouver, British Columbia, Canada.}

\begin{document}

\maketitle

%%%%%%%%%%%%%%%%%%%%%%%%%%%%%%%%%%%%%%%%%%%%%%%%%%%%%%%%%%%%%%%%%%

\begin{abstract}
\noindent Air pollution constitutes the highest environmental risk factor in relation to heath. In order to provide the evidence required for health impact analyses, to inform policy and to develop potential mitigation strategies comprehensive information is required on the state of air pollution. Information on air pollution traditionally comes from ground monitoring (GM) networks but these may not be able to provide sufficient coverage and may need to be supplemented with information from other sources (e.g. chemical transport models; CTMs). However, these may only be available on grids and may not capture micro-scale features that may be important in assessing air quality in areas of high population. We develop a model that allows calibration between multiple data sources available at different levels of support by allowing the coefficients of calibration equations to vary over space and time, enabling downscaling where the data is sufficient to support it. The model is used to produce high-resolution (1km $\times$ 1km) estimates of NO$_2$ and PM$_{2.5}$ across Western Europe for 2010-2016. Concentrations of both pollutants are decreasing during this period, however there remain large populations exposed to levels exceeding the WHO Air Quality Guidelines and thus air pollution remains a serious threat to health.
\end{abstract}

\newpage

%%%%%%%%%%%%%%%%%%%%%%%%%%%%%%%%%%%%%%%%%%%%%%%%%%%%%%%%%%%%%%%%%%
%%%%%%%%%%%%%%%%%%%%%%%%%%%%%%%%%%%%%%%%%%%%%%%%%%%%%%%%%%%%%%%%%%
%%%%%%%%%%%%%%%%%%%%%%%%%%%%%%%%%%%%%%%%%%%%%%%%%%%%%%%%%%%%%%%%%%

\section{Introduction}

%%%%%%%%%%%%%%%%%%%%%%%%%%%%%%%%%%%%%%%%%%%%%%%%%%%%%%%%%%%%%%%%%%
%%%%%%%%%%%%%%%%%%%%%%%%%%%%%%%%%%%%%%%%%%%%%%%%%%%%%%%%%%%%%%%%%%
%%%%%%%%%%%%%%%%%%%%%%%%%%%%%%%%%%%%%%%%%%%%%%%%%%%%%%%%%%%%%%%%%%

Ambient (outdoor) air pollution is a major cause of death and disease globally. The adverse health effects include both morbidity, such as increased hospital admissions and emergency room visits, as well as increased risk of mortality \citep{kampa2008human, hoek2013long}. In Europe, there is a long history of national and international regulatory approaches to air quality management. Starting with the Clean Air Act of 1956 in the United Kingdom, successive legislation, including the European emission standards passed in the European Union in 1992, has succeeded in reducing ambient air pollution levels over time \citep{shaddick2014case, turnock2016impact, guerreiro2014air, kuklinska2015air}. Current statutory limits state that annual average concentrations of NO$_{2}$ and PM$_{2.5}$ should not exceed 40 $\mu$g/m$^{3}$ and 25 $\mu$g/m$^{3}$ respectively \citep{EEAAQGs}. The World Health Organization (WHO) Air Quality Guidelines (AQG), that are designed to protect the public from the adverse health effects of air pollution, are currently the same as the EU statutory limits for NO$_{2}$ stating annual averages should not exceed 40 $\mu$g/m$^{3}$, but much lower than the the EU statutory limits for PM$_{2.5}$ stating annual averages should not exceed 10 $\mu$g/m$^{3}$ \citep{world2006air} .\\

\noindent The Directive on Ambient Air Quality and Cleaner Air for Europe requires air quality to be assessed throughout the territory of each member state. It requires that the fixed measurements should be used as a primary source of information for such assessment in the polluted areas \citep{union2008directive}. While this might be suitable for assessing adherence to statutory limits, in some cases ground measurements may not be able to provide sufficient spatial (and/or temporal) coverage. For example, in epidemiological studies measures of exposure required for each participant, often geo-located to their place of residence, and in estimating the country-level burden of disease, exposures are required for entire populations. In these cases, comprehensive sets of estimated levels of air pollution are required and although in Europe (and other high-income settings) there is extensive monitoring in urban areas there is considerably less in rural areas and the density of monitoring varies considerably. As a consequence, ground monitors alone cannot provide a complete picture of air pollution alone. \\

\noindent Comprehensive coverage of information, related to air quality over space and time can be obtained from other sources, such as remote sensing satellites (SAT) or chemical transport models (CTM). Examples of estimates from SAT include those produced by \citep{van2016global} with examples of a number of CTMs available include the GEOS-Chem \citep{bey2001global}, TM5-FASST \citep{van2014multi}, MOZART \citep{emmons2010description} and the Community Multiscale Air Quality (CMAQ) \citep{CMAQ} model. Here, we use the Monitoring Atmospheric Composition and Climate II (MACC-II) Ensemble model for Europe \citep{inness2013macc}, a model that combines estimates from seven separate regional CTMs. However, there are a number of issues using these estimates directly. The output of SATs and CTMs takes the form of gridded estimates and often have a low spatial resolution (e.g. 10km$\times$10km in the case of MACC-II CTM) and as such cannot capture micro-scale features that may be reflected in ground measurements. Therefore, it needs to be acknowledged that measurements from ground monitors and estimates from SAT or CTMs are fundamentally different quantities, with the latter is subject to uncertainties and biases arising from errors in inputs and possible model misspecification. In addition, the information they provide will be available at different geographical scales, i.e. point locations vs. grid-cells, an issue termed the `change of support problem' by \citet{gelfand2001change}, and a model will be required that can align the different sources in the spatial (and possibly temporal) domains. \\

\noindent One approach to linking data at different resolutions is to use spatially varying coefficient models, often referred to as \emph{downscaling} models. In a downscaling model, the parameters in the calibration equation are allowed to vary continuously over space (and potentially time) allowing predictions to be made at the point level thus allowing local, sub-grid-cell variation. Examples of downscaling in this setting include \citet{van2006statistical}, who modelled  PM$_{10}$ concentrations over Western Europe using  information from both satellite observations and a CTM; \citet{mcmillan2010combining} who modelled PM$_{2.5}$ in the North Eastern U.S. using estimates from the Community Multi-scale Air Quality (CMAQ) numerical model; \citep{van2016global} modelled annual average PM$_{2.5}$ calibrating ground measurements against estimates from both satellite remote sensing and CTMs; \citet{kloog2014new} who modelled PM$_{2.5}$ in the North-eastern U.S. using satellite-based aerosol optical depth (AOD); and \citet{berrocal2010spatio} and \citet{zidek2012combining} who modelled ozone in the Eastern U.S. (Eastern and Central in the case of Zidek {\em et al.})  using estimates from CMAQ and a variant of the MAQSIP (Multiscale Air Quality Simulation Platform) model respectively. \\

\noindent Here, we develop a model that allows calibration between data sources that are available at different levels of support, for example, ground monitors at point locations and estimates from  CTMs on grid-cells.  Set within a Bayesian hierarchical framework, the coefficients of calibration equations are allowed to vary continuously over space and time, enabling downscaling where ground monitoring data is sufficient to support it. We are specifically interested in the implementation of complex models for larger scale problems that may result in difficulties when attempting to use the methods for implementation proposed in the above examples, especially those using Markov Chain Monte Carlo. Here we use techniques that perform approximate Bayesian inference based on integrated nested Laplace approximations (INLA)  \citep{rue2009approximate} that allow high--resolution estimates of exposures to air pollution to be produced, together with associated measures of uncertainty. \\

\noindent The remainder of the paper is organised as follows: Section \ref{sec::data} provides a summary of the data used throughout this article.  In Section \ref{sec::genmod}, we introduce the proposed statistical models and the framework for performing inference, with Section \ref{sec::simstudy}  assessing the accuracy with which the model can estimate the parameters using simulated datasets. In Section \ref{sec::casestudy}, a case study is presented in which we examine the predictive ability of the model and produce high-resolution annual estimates of NO$_2$ and PM$_{2.5}$ across Western Europe for 2010-2016. Finally, Section \ref{sec::discuss} contains a discussion with concluding thoughts and suggestions for future work.

%%%%%%%%%%%%%%%%%%%%%%%%%%%%%%%%%%%%%%%%%%%%%%%%%%%%%%%%%%%%%%%%%%
%%%%%%%%%%%%%%%%%%%%%%%%%%%%%%%%%%%%%%%%%%%%%%%%%%%%%%%%%%%%%%%%%%

\section{Data} \label{sec::data}

%%%%%%%%%%%%%%%%%%%%%%%%%%%%%%%%%%%%%%%%%%%%%%%%%%%%%%%%%%%%%%%%%%
%%%%%%%%%%%%%%%%%%%%%%%%%%%%%%%%%%%%%%%%%%%%%%%%%%%%%%%%%%%%%%%%%%

The study region consists of 20 countries within Europe: Austria, Belgium, Denmark, Finland, France, Germany, Greece, Hungary, Ireland, Italy, Liechtenstein, Lithuania, Luxembourg, Netherlands, Norway, Portugal, Spain, Sweden, Switzerland and the United Kingdom. The sources of data used here can be allocated to one of four groups: (i) ground monitoring data; (ii) estimates from CTMs; (iii) other sources including land--use and topography and (iv) estimates of population counts. Ground monitoring is available at a distinct number of locations, whereas the latter three groups are available on grids and provide coverage of the entire study area with no missing data. \\

\noindent Annual average concentrations of NO$_2$ and PM$_{2.5}$ (measured in $\mu$g/m$^{3}$) from between 2010 and 2016 were extracted from the Air Quality e-Reporting database \citep{EEADatabase}. The database is maintained by the European Environment Agency (EEA) and provides a comprehensive and quality--assured source of information of air quality from all national and local monitoring networks in EU Member States and other participating countries. Measurements were used if they had $\geq$75\% of daily coverage over a year. \\

\noindent Numerically simulated estimates of NO$_2$ and PM$_{2.5}$ were obtained from the MACC-II Ensemble model available hourly on a 10km $\times$ 10km resolution grid at an hourly temporal resolution. In each grid-cell, the Ensemble model value was defined as the median value (in $\mu$g/m$^{3}$) of the following seven individual regional CTMs: CHIMERE from INERIS, EMEP from MET Norway, EURAD-IM from University of Cologne, LOTOS-EUROS from KNMI and TNO, MATCH from SMHI, MOCAGE from METEO-France and SILAM from FMI (see \citep{inness2013macc} and \citep{CAMS} for example). Estimates were aggregated over time to obtain annual average concentrations of NO$_2$ and PM$_{2.5}$ for each grid-cell.\\ 

\noindent Information on roads was extracted from the EuroStreets digital road network version 3.1. The road data was classified into 'all' and 'major' roads using the classification available in EuroStreets. These were then projected on to a 100m $\times$ 100m resolution grid, which were also aggregated to obtain estimates on a 1km $\times$ 1km resolution grid, with each cell the sum of road length within the cell. Information on land use was obtained using the European Corine Land Cover (ECLC) 2006 data set was obtained \citep{Corine2006}. This dataset covered the whole study area except Greece. For Greece, ECLC 2000 was used \citep{Corine2000}. Six classes (proportion of areas that are residential, industry, ports, urban green space, built up, natural land) were retrieved for use in modelling. These estimates were available on a high-resolution grid of 100m $\times$ 100m resolution and were also aggregated to obtain a 1km $\times$ 1km resolution grid. Information on altitude was obtained from the SRTM Digital Elevation Database version 4.1 with a resolution of approximately 90m resolution and aggregated to obtain 100m$\times$100m and 1km$\times$1km resolution grids \citep{SRTM90m}. SRTM data is only available up to 60$^{\degree}$N and was therefore supplemented with data from with Topo30 data. For more information, see \cite{de2016development}. A comprehensive set of population data on a high-resolution grid was obtained from Eurostat. The GEOSTAT 2011 version 2.0.1 database provides estimates of population at a 1km $\times$ 1km resolution across Europe in 2011 \citep{GEOSTAT}.

%%%%%%%%%%%%%%%%%%%%%%%%%%%%%%%%%%%%%%%%%%%%%%%%%%%%%%%%%%%%%%%%%%
%%%%%%%%%%%%%%%%%%%%%%%%%%%%%%%%%%%%%%%%%%%%%%%%%%%%%%%%%%%%%%%%%%

\section{Statistical Modelling} \label{sec::genmod}

%%%%%%%%%%%%%%%%%%%%%%%%%%%%%%%%%%%%%%%%%%%%%%%%%%%%%%%%%%%%%%%%%%
%%%%%%%%%%%%%%%%%%%%%%%%%%%%%%%%%%%%%%%%%%%%%%%%%%%%%%%%%%%%%%%%%%

\subsection{A spatio-temporal downscaling model for air pollution} \label{sec::mod}

%%%%%%%%%%%%%%%%%%%%%%%%%%%%%%%%%%%%%%%%%%%%%%%%%%%%%%%%%%%%%%%%%%

\noindent   Statistical calibration (or downscaling) is based on estimating relationships between measurements, $Y_{st}$ available at a discrete set of locations  $s \in S= \{s_1,\dots,s_{N_S}\}$ and time points $t\in T=\{t_1,\ldots, t_{N_T}\}$ and a set of covariates $X_{r\ell t}$, $r=1,\dots, R$, for example CTMs, satellite remote sensing, land use indicators, and topography, on grids of $N_{L_r}$ cells $\ell \in  \{\ell_{1},\ell_{2},\ldots, \ell_{N_{L_r}}\}$ at time $t$.\\

\noindent Our proposed model calibrates information from gridded covariates, $X_{r\ell t}$, against the ground measurements, $Y_{st}$, with both fixed and spatially and temporarily varying random effects for both intercepts and covariates,
\begin{equation}
\label{eqn:std}
	Y_{st} =  \tilde{\beta}_{0st} + \sum_{p \in P} \beta_{p}X_{p\ell_{ps}t} + \sum_{q \in Q}\tilde{\beta}_{qst}X_{q\ell_{qs}t} + \epsilon_{st}
\end{equation}
where $\epsilon_{st} \sim^{iid} N(0,\sigma^2_\epsilon)$. Here, $\ell_{ps}$ denotes the grid-cell in grid $p$ containing the point location $s$. The set of $R$ covariates contains two groups, $R = (P,Q)$, where $P$ have fixed effects, $\beta_{q}$, and $Q$ are assigned spatio--temporally varying random effects, $\tilde{\beta}_{pst}$.\\

\noindent The spatio--temporally varying coefficients $\tilde{\beta}_{pst}$, $p=0,1,\ldots,P$,  comprise of fixed and varying effects 
\begin{eqnarray*}
	\tilde{\beta}_{pst} = \beta_p + \beta_{pst},
\end{eqnarray*} 
where $\beta_{pst}$ provides temporal and spatial adjustments around $\beta_p$, a fixed effect. For clarity of exposition, the following description is restricted to the coefficients associated with a single covariate, that is $\tilde{\beta}_{st}$. In time, $\boldsymbol{\beta}_t = (\beta_{1t}, \beta_{2t},\ldots, \beta_{N_St})$ is assumed to evolve as a first-order autoregressive process 
\begin{equation*}
	\boldsymbol{\beta}_{t} = \rho  \boldsymbol{\beta}_{t-1} +  \boldsymbol{\omega}_{t}
\end{equation*} 
where $\boldsymbol{\omega}_t = (\omega_{1t}, \omega_{2t},\ldots, \omega_{N_St})$ are assumed to be independent and identically distributed draws from a stationary, isotropic, zero-mean Gaussian random field, $N(0, \sigma^2_\omega\Sigma)$, with Mat\'{e}rn covariance function 
\begin{equation}
	Cov(\omega_{s_it_i},\omega_{s_jt_j}) = \delta_{ij}\frac{\sigma_{\omega}^2}{2^{\nu-1} \Gamma(\nu)}(\kappa \|s_i-s_j\|)^\nu K_\nu(\kappa \|s_i-s_j\|) \label{eqn::matern}
\end{equation}
\noindent where $\delta_{ij}$ is the Dirac delta function, $K_\nu$ is the modified Bessel of the second kind, $\sigma_\omega^2$ is the overall variance, $\nu$ controls the smoothness of the spatial process, $\kappa$ controls the strength of the distance/correlation relationship and $\|\cdot\|$ is Euclidean distance. Moreover, $\omega_{st}$ is modelled as being separable over space and time with the covariance structure is constructed using a Kronecker product \citep{cameletti2013spatio}. \\

\noindent Gaussian priors, $N(0, 1000)$, are assigned to each of the fixed effects $\beta_0$, $\beta_p$ and $\beta_q$. For the spatio-temporal random effects, the smoothness parameter is fixed, $\nu = 1$ as in \citet{lindgren2011spde}. Penalised Complexity (PC) priors are used for the variance of the observations, $\sigma^2_\epsilon,  \mathbb{P}(\sigma_{\epsilon} > 1) = 0.1$, the range parameter ($\mathbb{P}(\kappa_p < 0.1) = 0.1$), standard deviation ($\mathbb{P}(\sigma_{\omega p} > 1) = 0.1$) and the autocorrelation parameter ($\mathbb{P}(\rho_{p} > 0) = 0.9$) \citep{simpson2017penalising,fuglstad2018constructing}. 

%%%%%%%%%%%%%%%%%%%%%%%%%%%%%%%%%%%%%%%%%%%%%%%%%%%%%%%%%%%%%%%%%%
%%%%%%%%%%%%%%%%%%%%%%%%%%%%%%%%%%%%%%%%%%%%%%%%%%%%%%%%%%%%%%%%%%

\subsection{Approximating the continuous spatio-temporal field}

%%%%%%%%%%%%%%%%%%%%%%%%%%%%%%%%%%%%%%%%%%%%%%%%%%%%%%%%%%%%%%%%%%
%%%%%%%%%%%%%%%%%%%%%%%%%%%%%%%%%%%%%%%%%%%%%%%%%%%%%%%%%%%%%%%%%%

In typical downscaling applications, the spatial processes governing the coefficients are modelled as Gaussian random fields (GRFs). The field is defined by the covariance matrix and performing inference with a large number of monitoring locations over many time points may be computational challenging using traditional methods of performing Bayesian inference (e.g. MCMC) as large, dense, linear systems must be solved at each iteration. This poor computational scaling with the size of the data is known as the `big N' problem. A number of methods specifically tailored to scaling up inference in spatial and spatio-temporal problems have been proposed over the past decade, and some of the most broadly used methods use a specially constructed finite-dimensional Gaussian random field that trades off scalability with accuracy \citep{cressie2008fixed, katzfuss2017multi, lindgren2011spde}.  A review of recent methods, as well as information about their performance on a simple spatial model, can be found in \citet{heaton2017methods}. \\ 

\noindent Here, we propose representing the continuous field by a approximation based on an (irregular) triangulation. A triangulation is a partition of a domain of interest into a collection of connected non-overlapping triangles. The triangles of a triangulation are formed by points given in the domain of interest. This will allow us to control the smoothness of the process, by allowing it to vary more quickly over space where there is data to support it and less so when there is less data, and therefore focussing the computational effort and thus enabling downscaling where the data is sufficient to support it. In the case of air pollution this may be more appropriate, as this allows us to define a more dense set of triangles where data are dense spatially and less where is it is more sparse. Furthermore, we use Delauney triangulation which places constraints on the maximum angle size and the triangulation to be convex \citep{hjelle2006triangulations}.\\

\noindent Once a triangulation of the domain has been defined, the spatial field can be approximated using, 
\begin{equation}
	\omega_{st} = \sum_{k=1}^{n} \phi_{ks}w_{kt} \label{eqn::approx}
\end{equation}
where $n$ is the number of vertices (or nodes) of the triangulation, $\{\phi_{ks}\}$ are a set of piecewise linear basis functions that are one at vertex $k$ and zero at all other vertices, and $\{w_{kt}\}$ are a set of stochastic weights. The weights $w_{kt}$ are assigned a zero-mean multivariate Gaussian distribution $\boldsymbol{w}_t \stackrel{\text{iid}}{\sim} N(0,\Sigma)$. \\

\noindent Here, we follow \cite{lindgren2011spde} and select the distribution of the weights such that we approximate the GRF by a Gaussian Markov random field (GMRF). A GMRF is a discretely indexed GRF, $\boldsymbol{Z}\sim N(0,\Sigma)$ and if one can be found that best represents our GRF then we will be able to take advantage of efficient computation, as typically, the inverse of the covariance matrix, $Q=\Sigma^{-1}$ will be sparse, due to a conditional independence property, in which $Z_i \ci  Z_j | \boldsymbol{Z}_{-ij} \iff \mathit{Q}_{ij}=0$ (where $\boldsymbol{Z}_{-ij}$ denotes $\boldsymbol{Z}$ with the $i^{th}$ and $j^{th}$ elements removed) \citep{rue2005gaussian}. The structure of the precision matrix of this process is defined by the triangulation of the domain and the set of bases functions used. In order to ensure a the Markov structure that is required for a GMRF, the set of basis functions should be piecewise linear,
\begin{equation*}
\phi_{ks} = 
\begin{cases}
1 & \text{  at vertex $k$ } \\
0 & \text{  at all other vertices}
\end{cases}.
\end{equation*}
\noindent then $\boldsymbol{w}_t \stackrel{\text{iid}}{\sim} N(0,\Sigma)$ will be a GMRF.

\noindent Furthermore, if the GRF, $\{\omega_s \; | \; s \in \mathbb{R}^d\}$, is assumed to have a Mat\'ern covariance function as in Equation (\ref{eqn::matern}), then the approximation given by Equation (\ref{eqn::approx}) is a finite element method solution to the following SPDE, 
\begin{equation*}
(\kappa^2 - \Delta)^{\alpha/2}(\tau\omega_s) = \mathcal{W}_s  \;\; \boldsymbol{s} \in \mathbb{R}^d, \;\; \alpha = \nu + d/2, \;\; \kappa>0,\;\; \nu > 0
\end{equation*}
where $(\kappa^2 - \Delta)^{\alpha/2}$ is a pseudo-differential operator, $\Delta$ is the Laplacian, $\kappa$ is the scale parameter, $\tau$ controls the variance and $\mathcal{W}_s$ is spatial white noise with unit variance. Any GRF model defined with a Mat\'ern covariance structure can be approximated by a GMRF in this way provided $\nu + d/2$ is integer valued. This approach can be extended to GRFs on manifolds, non-stationary and anisotropic covariance structures \citep{lindgren2011spde,bakka2018spatial,krainski2018advanced}. 

%%%%%%%%%%%%%%%%%%%%%%%%%%%%%%%%%%%%%%%%%%%%%%%%%%%%%%%%%%%%%%%%%%
%%%%%%%%%%%%%%%%%%%%%%%%%%%%%%%%%%%%%%%%%%%%%%%%%%%%%%%%%%%%%%%%%%

\subsection{Inference}

%%%%%%%%%%%%%%%%%%%%%%%%%%%%%%%%%%%%%%%%%%%%%%%%%%%%%%%%%%%%%%%%%%
%%%%%%%%%%%%%%%%%%%%%%%%%%%%%%%%%%%%%%%%%%%%%%%%%%%%%%%%%%%%%%%%%%

The model presented in Section \ref{sec::mod} can be expressed in general form as 
\begin{eqnarray} 
	\eta_i = \mathbb{E}(Y_{st}) &=& \beta_0 + \sum_{p=1}^{P} \beta_p X_{pst} + \sum^Q_{q=1} f_q(X_{q\ell t}) ,	
	\label{eqn::LGM}
\end{eqnarray}
where, $\beta_0$ is an overall intercept term, the set of $\beta_p$ $(p=1,\dots, P)$ are the coefficients associated with covariates $X_{qst}$ that are not misaligned, the set of functions, $f_1(\cdot), \ldots, f_P(\cdot)$ represent the random effects which can take the form of random intercepts and slopes, non-linear effects of covariates, spatial, temporal and spatio-temporal random effects.  \\

\noindent  All unknown parameters are collected into two sets: $\boldsymbol{\theta}= (\beta_q, f_q)$, which contains all of the parameters, and $\boldsymbol{\psi}$, which contains all of the hyperparameters that control the variability and strength of the relationships between the observations and the fixed and random effects. In the model presented in Section \ref{sec::genmod}, the parameter field $\boldsymbol{\theta}$, contains the parameters of the calibration equation and will include the fixed effects, $\beta_p$ and the spatio-temporal random effects $\beta_{pst}$. The set $\boldsymbol{\psi}$ contains the Gaussian noise parameter $\sigma^2_\epsilon$, along with  the variance, range parameter and autocorrelation parameter associated with the spatio-temporal random effects, $\sigma^2_{\omega p}$, $\kappa_{p}$ and $\rho_{p}$. Assigning a Gaussian distribution to the set of parameters $\boldsymbol{\theta} \sim N(0,\Sigma(\boldsymbol{\psi}_2))$ results in a latent Gaussian model (LGM). LGMs are a subclass of models such as generalized linear or additive (mixed) models, temporal, spatial or spatio-temporal models \citep{rue2009approximate}. \\\\

\noindent The aim is to produce a set of high-resolution exposures to air pollution over an entire study area in time. The marginal posterior distribution for a prediction in a particular location, $s$, and time, $t$, can be expressed as
\begin{equation}
	p(\hat{Y}_{st}|\boldsymbol{Y}) = \int \int p(\hat{Y}_{st}| \boldsymbol{\theta},\boldsymbol{\psi},\boldsymbol{Y})p(\boldsymbol{\theta}|\boldsymbol{\psi},\boldsymbol{Y})p(\boldsymbol{\psi}|\boldsymbol{Y})d\boldsymbol{\theta} d\boldsymbol{\psi}
	\label{eqn:Prediction}
\end{equation}
\noindent There is also interest in finding the marginal posterior densities for each $\psi_k$ and $\theta_j$ given the observed data $\boldsymbol{Y}$, 
\begin{equation}
	\begin{split}
		p(\psi_k|\boldsymbol{Y}) &= \int p(\boldsymbol{\psi}|\boldsymbol{Y}) d\boldsymbol{\psi}_{-k} \\
		p(\theta_j|\boldsymbol{Y}) &= \int p(\theta_j|\boldsymbol{\psi},\boldsymbol{Y})p(\boldsymbol{\psi}|\boldsymbol{Y})d\boldsymbol{\psi}.
		\end{split} \label{eqn::inla}
\end{equation}
Here, $\boldsymbol{\psi}_{-k}$ denotes the set of parameters, $\boldsymbol{\psi}$, with the $k^{th}$ entry removed. In most cases, these will not be analytically tractable and therefore approximations of the posterior distributions $p(\boldsymbol{\psi}|\boldsymbol{Y})$ and $p(\theta_j|\boldsymbol{\psi},\boldsymbol{Y})$ as well as numerical integration will be needed. \\

\noindent The approximation for the joint posterior, $p(\boldsymbol{\psi}|\boldsymbol{Y})$, is given by
\begin{equation*}
	\tilde{p}(\boldsymbol{\psi}|\boldsymbol{Y}) \propto \left.\frac{p(\boldsymbol{Y},\boldsymbol{\theta}, \boldsymbol{\psi})}{\tilde{p}(\boldsymbol{\theta}|\boldsymbol{\psi}, \boldsymbol{Y})} \right| _{\boldsymbol{\theta} = \hat{\boldsymbol{\theta}}(\boldsymbol{\psi})} 
\end{equation*}
where $\tilde{p}(\boldsymbol{\theta}|\boldsymbol{\psi}, \boldsymbol{Y})$ is a Gaussian approximation of $p(\boldsymbol{\theta}|\boldsymbol{\psi}, \boldsymbol{Y})$ evaluated at the mode $\hat{\boldsymbol{\theta}}(\boldsymbol{\psi})$ of the distribution $\boldsymbol{\theta}|\boldsymbol{\psi}$. The approximation, $\tilde{p}(\boldsymbol{\psi}|\boldsymbol{Y})$, is equivalent to a Laplace approximation, and it is exact if $\tilde{p}(\boldsymbol{\theta}|\boldsymbol{\psi}, \boldsymbol{Y})$ is Gaussian. The approximation used for the posterior, $p(\theta_j|\boldsymbol{\psi},\boldsymbol{Y})$ is given by
\begin{equation*}
\tilde{p}(\theta_j|\boldsymbol{\psi},\boldsymbol{Y}) \propto   \left.\frac{p(\boldsymbol{Y},\boldsymbol{\theta}, \boldsymbol{\psi})}{\tilde{p}(\boldsymbol{\theta}_{-j}| \theta_{j}, \boldsymbol{\psi}, \boldsymbol{Y})} \right| _{\boldsymbol{\theta}_{-j} = \hat{\boldsymbol{\theta}}_{-j}(\theta_{j}, \boldsymbol{\psi})} 
\end{equation*}
where $\tilde{p}(\boldsymbol{\theta}_{-j}| \theta_{j}, \boldsymbol{\psi}, \boldsymbol{Y})$ is a Gaussian approximation of the distribution $p(\boldsymbol{\theta}_{-j}| \theta_{j}, \boldsymbol{\psi}, \boldsymbol{Y})$. The distribution $\tilde{p}(\theta_j|\boldsymbol{\psi},\boldsymbol{Y})$ is then obtained by taking Taylor expansions of $p(\boldsymbol{Y},\boldsymbol{\theta}, \boldsymbol{\psi})$ and $\tilde{p}(\boldsymbol{\theta}_{-j}| \theta_{j}, \boldsymbol{\psi}, \boldsymbol{Y})$, up to third order, aiming to correct a Gaussian approximation for location errors due to potential skewness \citep{rue2009approximate}. \\

\noindent In order to estimate the marginal posterior distributions given by Equation (\ref{eqn::inla}), a set of integration points and weights are built using the distribution $\tilde{p}(\boldsymbol{\psi}|\boldsymbol{Y})$. Firstly, the mode of $\tilde{p}(\boldsymbol{\psi}|\boldsymbol{Y})$ is found numerically by Newton-type algorithms. Around the mode, the distribution $\log(\tilde{p}(\boldsymbol{\psi}|\boldsymbol{Y}))$ is evaluated over a grid of $K$ points $\{\boldsymbol{\psi}^{(k)}\}$, each with associated integration weights $\{\Delta^{(k)}\}$. If the points define a regular lattice, then the integration weights will be equal. The marginal posteriors, $p(\psi_k|\boldsymbol{Y})$, are obtained using numerical integration of an interpolant of $\log(\tilde{p}(\boldsymbol{\psi}|\boldsymbol{Y}))$. The marginal posteriors, $p(\theta_j|\boldsymbol{Y})$, are obtained using numerical integration
\begin{equation}
\label{eqn:NI}
	p(\theta_j|\boldsymbol{Y}) = \sum_{k} \tilde{p}(\theta_j|\boldsymbol{\psi}^{(k)},\boldsymbol{Y}) \tilde{p}(\boldsymbol{\psi}^{(k)}|\boldsymbol{Y})\Delta^{(k)} 
\end{equation}
where $\tilde{p}(\theta_j|\boldsymbol{\psi}^{(k)},\boldsymbol{Y})$ and $\tilde{p}(\boldsymbol{\psi}^{(k)}|\boldsymbol{Y})$ are the posterior distributions $\tilde{p}(\boldsymbol{\psi}|\boldsymbol{Y})$ and $\tilde{p}(\theta_j|\boldsymbol{\psi},\boldsymbol{Y})$ evaluated at the set of integration points $\{\boldsymbol{\psi}^{(k)}\}$ while $\Delta^{(k)}$ are integration weights \citep{martins2013bayesian}.  \\

\noindent Inference of the model presented in Section \ref{sec::mod} can be implemented using R-INLA \citep{rue2012r}. For further details, see \citep{rue2009approximate}.

%%%%%%%%%%%%%%%%%%%%%%%%%%%%%%%%%%%%%%%%%%%%%%%%%%%%%%%%%%%%%%%%%%
%%%%%%%%%%%%%%%%%%%%%%%%%%%%%%%%%%%%%%%%%%%%%%%%%%%%%%%%%%%%%%%%%%

\subsection{Prediction}

%%%%%%%%%%%%%%%%%%%%%%%%%%%%%%%%%%%%%%%%%%%%%%%%%%%%%%%%%%%%%%%%%%
%%%%%%%%%%%%%%%%%%%%%%%%%%%%%%%%%%%%%%%%%%%%%%%%%%%%%%%%%%%%%%%%%%

In a fully Bayesian analysis, predictions (as given by Equation (\ref{eqn:Prediction})) are treated as unknown parameters and posterior distributions of these quantities are estimated alongside the model parameters. This may cause computational issues particularly when predicting at a very large number of locations in time, again due to the need to manipulate large covariance matrices. \\

\noindent One approach would be to perform inference with predictions at a small number of locations simultaneously and to repeat this a number of times to obtain a full set of predictions, as used in \citet{Shaddick2017DIMAQ} when predicting global air quality on a high resolution grid ($0.1^{\degree}$ $\times$ $0.1^{\degree}$ resolution). However, uncertainty may be underrepresented as the joint variance between predictions is ignored. An alternative approach, which is the one taken here, is to take joint samples from the posterior distributions, $\tilde{p}(\boldsymbol{\theta}|\boldsymbol{\psi},\boldsymbol{Y})$ and $p(\boldsymbol{\psi}|\boldsymbol{Y})$ and use each set of samples (of the model coefficients) to create predictions using Equation (\ref{eqn:std}), resulting in a set of joint predictions of the quantity of interest. This provides an efficient method for predicting at any required location in space and time as once the samples of the model coefficients are obtained, prediction is done using a linear combination. Summarising the joint samples at each location produces marginal predictive posterior distributions. Furthermore, full posterior distributions for other quantities of interest such as annual average concentrations, population-weighted annual averages concentrations or changes over time can be produced. \\

 \noindent Although R-INLA estimates marginal posterior densities, as shown in Equation (\ref{eqn::inla}), it is possible to sample from the joint posterior distribution using the function \texttt{inla.posterior.sample} in the R-INLA package. In computing the approximation to the required distributions as shown in Equation (\ref{eqn:NI}) the approximated distributions at the integration points can be retained. Joint samples from the posteriors can be obtained by sampling from Gaussian approximations at the integration points for all of the parameters.  

%%%%%%%%%%%%%%%%%%%%%%%%%%%%%%%%%%%%%%%%%%%%%%%%%%%%%%%%%%%%%%%%%%
%%%%%%%%%%%%%%%%%%%%%%%%%%%%%%%%%%%%%%%%%%%%%%%%%%%%%%%%%%%%%%%%%%
%%%%%%%%%%%%%%%%%%%%%%%%%%%%%%%%%%%%%%%%%%%%%%%%%%%%%%%%%%%%%%%%%%

\section{Simulation Study}
\label{sec::simstudy}

%%%%%%%%%%%%%%%%%%%%%%%%%%%%%%%%%%%%%%%%%%%%%%%%%%%%%%%%%%%%%%%%%%
%%%%%%%%%%%%%%%%%%%%%%%%%%%%%%%%%%%%%%%%%%%%%%%%%%%%%%%%%%%%%%%%%%
%%%%%%%%%%%%%%%%%%%%%%%%%%%%%%%%%%%%%%%%%%%%%%%%%%%%%%%%%%%%%%%%%%

\begin{figure}[H]
	\centering
	\includegraphics[width = 0.6\linewidth, clip = TRUE, trim = {4cm 7cm 4cm 5cm}]{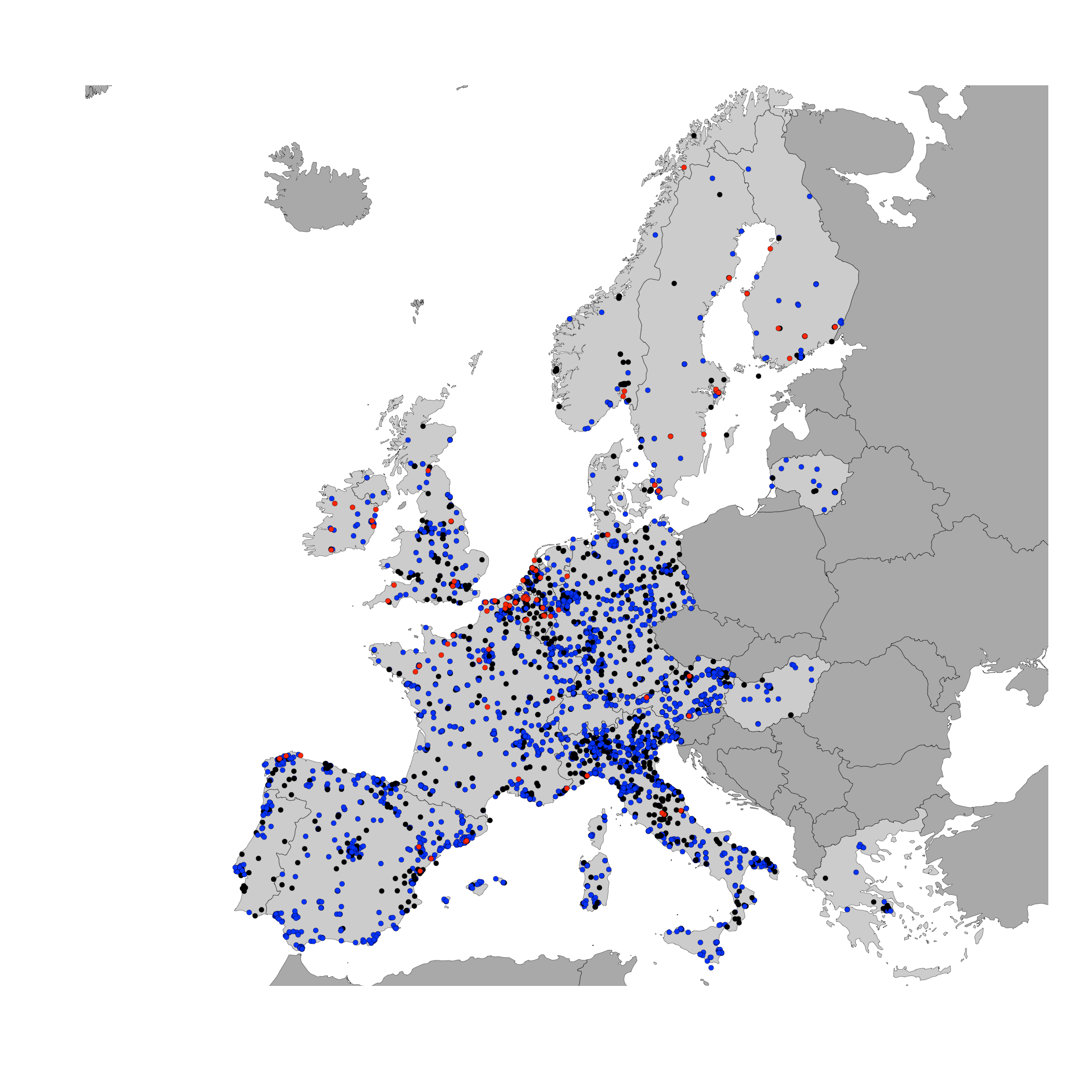}
	\caption{Locations of ground monitors measuring NO$_2$ or PM$_{2.5}$ between 2010 and 2016 in the study area. Red dots denote monitors that only measure PM$_{2.5}$ during this period, with blue dots denoting monitors only measuring NO$_2$ and black dots denoting monitors measuring both and PM$_{2.5}$.} \label{fig::mapgm}
\end{figure}

 In this section we present a simulation study that assesses the accuracy with which the model can estimate the parameters. We fit a series of models to multiple simulated datasets, the basis of which are the estimates from the MACC-II Ensemble CTM (see Sections \ref{sec::data} and \ref{sec::casestudy}), from which measurements are generated at the locations of the NO$_2$ monitoring sites in Europe (as seen in Figure \ref{fig::mapgm}). For computational ease, data is only generated for one time point. The study region is a 400 by 700  grid containing 280,000  cells, $X_l: l=1,...N_L$, each  of which is assigned the square root of the estimated annual average concentration of NO$_2$ from the CTM for 2016. Table \ref{tab::DistGMs} shows the distribution of the number of monitors within grid-cells. Of the grid-cells that contained ground monitors, 26.3\% contained more than one and will be used assess to the ability of the model to capture within grid-cell variability.\\
 
\noindent When generating the data, we assume that there is spatial structure on both the intercept term and the coefficient representing the association between ground measurements and estimates from the CTM. We show the efficacy of the modelling approach shown in Section \ref{sec::genmod} compared to simpler alternatives including linear regression and a model in which only the intercept is allowed to vary spatially (akin to a kriging model). \\

\begin{table}[t]
	\centering
	\caption{Distribution of the number of ground monitors within CTM grid-cells.}
	\label{tab::DistGMs}
	\begin{tabular} {l | c c c c c c c c}
		Number of GMs & 1 & 2 & 3 & 4 & 5 & 6 & 7 & 8+ \\
		\hline 
		Number of grid-cells & 1258 & 270  & 98 &  40  & 16 & 10 & 6 & 9
	\end{tabular}
\end{table}

\noindent Square roots of ground measurements at location $s$, $Y_s$, are generated from the following model:
\begin{equation}
	Y_s = \tilde{\beta}_{0s} + \tilde{\beta}_{1s}X_{l_s} + \epsilon_s \label{eqn::simmodel}
\end{equation}
where $\epsilon_s \sim N(0, \sigma^2_\epsilon)$. The coefficients, $\tilde{\beta}_{0s}$ and $\tilde{\beta}_{1s}$, are as follows
\begin{eqnarray*}
	\begin{split}
		\tilde{\beta}_{0s} &=& \beta_{0} + \beta_{0s}\\
		\tilde{\beta}_{1s} &=& \beta_{1} + \beta_{0s}.
	\end{split}
\end{eqnarray*}
where $\beta_{0}$ is a fixed effect representing the overall mean (level of measured NO$_2$) and $\beta_{0s}$ is a spatial random effect from the Mat\'ern class of spatial covariance functions, given by Equation (\ref{eqn::matern}). Similarly, $\beta_{1}$ is a fixed effect representing the association between the estimates from the CTM and measurements of measured NO$_2$ with $\beta_{1s}$ a spatial random effect with the same form as for $\beta_{0s}$. The variances of the spatial processes for the intercept and association between the CTM and GM are $\sigma^2_0$ and $\sigma^2_1$  respectively, with the range (of the spatial processes) being $\kappa_0$ and $\kappa_1$. \\

\noindent Table \ref{tab::Results} shows the values of the parameters of the models used to produce ten simulated datasets in the simulation study. We fit a series of three models representing different levels of complexity:
\begin{itemize}
	\item[\textbf{(i)}] The intercept and the coefficient associated with the CTM are assumed constant over space,
	\item[\textbf{(ii)}] A model with spatial structure on the intercept only with the coefficient associated with the CTM is constant over space,
	\item[\textbf{(iii)}] A model with spatial structure on the intercept and coefficient associated with CTM (as described in Section \ref{sec::mod}). 
\end{itemize}
These models were run on each simulated dataset, with each marginal posterior distribution associated with each parameter extracted. Table \ref{tab::Results} also show the results of the simulation study in the form of the median of the estimates from each simulated dataset along with the median biases associated with the estimates. \\

\noindent In the simulated data, there is spatial variation in both the intercept and the coefficient associated with the CTM, therefore it can be seen that fitting a model with no spatial variation in either coefficient (Model (i)) results in considerable over estimation of the random error component, $\sigma^2_\epsilon$, due to the need to incorporate (structured) variation in the data (observed via $\beta_{0s}$ and $\beta_{1s}$). Adding spatial variation in the intercept term (Model (ii)) sees a substantial improvement in the estimation of  $\sigma^2_\epsilon$ but the parameters associated with the spatial process for the intercept ($\kappa_0$  and $\sigma^2_0$) are both overestimated due to the use of a single spatial process to accommodate the spatial variation in both the intercept and the association with CTM. A model which also adds spatial structure to the coefficient associated with the CTM (Model (iii)) shows the lowest bias for all parameters, with notable improvements compared to Model (ii) when estimating the spatial variability in the intercept now that variation can be correctly assigned to the coefficient associated with CTM. 

\begin{table}[t]
	\centering
	\caption{Summaries of results from fitting three models to ten simulated datasets: median estimated value and bias for given parameters. Results are the median values from fitting the models to multiple simulated datasets.}
	\label{tab::Results}
	\begin{tabular} {c lllllll}
		\hline \\[-8pt]
		          &            & \multicolumn{2}{c}{\textbf{Model (i)}}& \multicolumn{2}{c}{\textbf{Model (ii)}} & \multicolumn{2}{c}{\textbf{Model (iii)}} \\
		       
		\cmidrule(lr){3-4}
	 	\cmidrule(lr){5-6}
	 	\cmidrule(lr){7-8}
		Parameter & True value & Estimate &  Bias & Estimate &  Bias & Estimate &  Bias \\[3pt]
		\hline \\[-5pt]
		$\beta_0$           & 1    & 0.6634 & -0.3366 &  0.7990 & -0.2009 &  0.9954 & -0.0045    \\
		$\kappa_0$          & 10   &  -     & -       & 22.7204 & 12.7204 &  9.7764 & -0.2235            \\
		$\sigma^2_0$        & 0.2  &  -     & - 	      &  0.3243 &  0.1243 &  0.1848 & -0.0152        \\[5pt]
		$\beta_1$           & 0.75 & 0.8910 &  0.1410 &  0.8052 &  0.0552 &  0.7565 &  0.0065 \\
		$\kappa_1$          & 15   &  -     & - 	      &   -     & - 	      & 15.2148 &  0.2149 \\
		$\sigma^2_1$        & 0.01 &  -     & -       &   -     & -       &  0.0242 &  0.0142 \\[5pt]
		$\sigma^2_\epsilon$ & 0.1  & 0.5846 &  0.4846 &  0.1014 &  0.0014 &  0.1001 &  0.0001 \\[5pt]
		\hline
	\end{tabular}
\end{table}

%%%%%%%%%%%%%%%%%%%%%%%%%%%%%%%%%%%%%%%%%%%%%%%%%%%%%%%%%%%%%%%%%%
%%%%%%%%%%%%%%%%%%%%%%%%%%%%%%%%%%%%%%%%%%%%%%%%%%%%%%%%%%%%%%%%%%
%%%%%%%%%%%%%%%%%%%%%%%%%%%%%%%%%%%%%%%%%%%%%%%%%%%%%%%%%%%%%%%%%%

\section{Case Study: Air Pollution in Western Europe} \label{sec::casestudy}

%%%%%%%%%%%%%%%%%%%%%%%%%%%%%%%%%%%%%%%%%%%%%%%%%%%%%%%%%%%%%%%%%%
%%%%%%%%%%%%%%%%%%%%%%%%%%%%%%%%%%%%%%%%%%%%%%%%%%%%%%%%%%%%%%%%%%
%%%%%%%%%%%%%%%%%%%%%%%%%%%%%%%%%%%%%%%%%%%%%%%%%%%%%%%%%%%%%%%%%%

The locations of 3436 ground monitoring sites active between 2010 and 2016 can be seen in Figure \ref{fig::mapgm} with summaries of the annual average concentrations of NO$_2$ and PM$_{2.5}$ by year shown in Table \ref{tab::summarygm}. The majority of the ground monitors are located in urban and highly-populated areas, with NO$_2$ being more extensively monitored than PM$_{2.5}$ (although  ground monitoring of PM$_{2.5}$ is increasing during this period). Annual concentrations of both  NO$_2$ and PM$_{2.5}$ are decreasing during this period, from a median of 25.0 $\mu$g/m$^{3}$ in 2016 to 20.4 $\mu$g/m$^{3}$ (NO$_2$) and from 15.8 $\mu$g/m$^{3}$ to 12 $\mu$g/m$^{3}$ (PM$_{2.5}$). \\

\begin{table}[t]
\footnotesize
\centering
	\caption{Summary of NO$_2$ and PM$_{2.5}$ concentrations (measured in $\mu$g/m$^{3}$) at ground monitoring sites, by year.}\label{tab::summarygm}
	\begin{tabular}{clc ccc ccc}	
	\hline \\[-8pt]
	\multicolumn{2}{l}{\begin{tabular}[c]{@{}l@{}} \textbf{Pollutant}, \\ \hspace{3mm}   \textbf{Summary}\\[3pt]\end{tabular}}  & \textbf{N} & \textbf{Mean} & \textbf{Std. Dev.} & \textbf{Median} & \textbf{Quantiles (2.5, 97.5\%)} \\[3pt] 
	\hline \\[-5pt]
	  \multicolumn{2}{l}{NO$_{2}$} &&&&\\
		  & 2010 & 2431 & 26.9 & 14.6 & 25.0 & (4.4, 61.1)\\
		  & 2011 & 2421 & 25.8 & 14.7 & 23.7 & (4.3, 61.3)\\ 
		  & 2012 & 2346 & 24.9 & 14.0 & 22.6 & (4.2, 59.1)\\ 
		  & 2013 & 2271 & 24.0 & 13.6 & 22.2 & (3.8, 57.0)\\ 
		  & 2014 & 2415 & 22.6 & 13.0 & 20.8 & (3.7, 53.2)\\ 
		  & 2015 & 2396 & 23.3 & 13.4 & 21.1 & (3.7, 54.5)\\
		  & 2016 & 2512 & 22.4 & 12.6 & 20.4 & (3.7, 51.5)\\[5pt]
		\multicolumn{2}{l}{PM$_{2.5}$} &&&&\\ 
		  & 2010 & 560 & 15.8 & 5.1 & 15.8 & (6.5, 26.1) \\
		  & 2011 & 667 & 16.2 & 5.7 & 16.1 & (6.4, 30.5)\\ 
		  & 2012 & 715 & 14.5 & 5.1 & 14.2 & (6.1, 27.7)\\ 
		  & 2013 & 796 & 14.3 & 4.8 & 14.3 & (5.5, 26.6)\\ 
		  & 2014 & 873 & 13.1 & 4.1 & 13.0 & (5.8, 22.5)\\ 
		  & 2015 & 928 & 13.4 & 4.9 & 12.9 & (5.6, 26.6)\\ 
		  & 2016 & 977 & 12.2 & 4.4 & 11.9 & (4.9, 24.0)\\[5pt]
	\hline
	\end{tabular}
\end{table}

\noindent Based on the findings from the simulation study, we fit a model with random effects on both the intercept terms and coefficients associated with the MACC-II Ensemble CTM (Model (iii) from Section \ref{sec::simstudy}). We utilise the data which is available for multiple years by extending the model to allow spatio-temporally variation relationships between GMs and CTM using the formulation of the model given in Equation (\ref{eqn:std}), with the relationship between the information on roads (length of all and major roads), land use (the proportion of areas that are residential, industry, ports, urban green space, built up, natural land), altitude assumed to be (fixed and) linear, with transformations of some of the covariates, notably altitude ($\sqrt{a/max(a)}$ where $a = \mbox{altitude} - min(\mbox{altitude})$), to address nonlinearity. See Section \ref{sec::data} for details on the data used. To allow for the skew in the ground measurements and the constraint of non-negativity, the square-root of the measurements are used for NO$_2$ and the (natural) logarithm of the measurements are used for PM$_{2.5}$. \\

\noindent Using cross-validation, we compare the model with spatio-temporal random effects against Models (i) and (iii) from Section \ref{sec::simstudy}). The three models considered are:
\begin{itemize}
	\item[\textbf{(i)}] {\bf Fixed effects} - The intercept and the coefficients associated with all covariates are assumed constant over space and time
	\item[\textbf{(iii)}] {\bf Spatial random effects} - A model with spatial structure on the intercept and coefficient associated with CTM, with all other coefficients associated with all covariates are assumed constant over space and time. 
	\item[\textbf{(iv)}] {\bf Spatio-temporal random effects}. A model with spatio-temporal structure on the intercept and coefficient associated with CTM, with all other coefficients associated with all covariates are assumed constant over space and time.
\end{itemize}

\noindent Cross-validation was performed by fitting the model to 25 training sets (each containing 80\% of the data) and comparing predictions with measurements from validation sets (the remaining 20\% of the data). Validation sets were obtained by taking a stratified random sample, based on classifying sites in terms of site type, country and year. The following metrics were calculated for each training--evaluation set combination:  R$^2$, root-mean-squared error (RMSE), and population-weighted root-mean-squared error (PwRMSE), Each is calculated for both within-sample evaluation (model fit) and out-of-sample evaluation (predictive accuracy). \\

\noindent The results of fitting the three models can be found in Table \ref{tab::evaluation} (See Appendix Table \ref{tab::evaluationbyyear} for results by year). It can be seen that using either of the downscaling models (Models (ii) and (iv)) provides a marked improvement in all metrics for both within-sample model fit and out-of-sample predictive ability when compared with the linear model (Model (i)). For example, using Model (ii) which contains only spatial variation in the intercept and the coefficient associated with the CTM, results in the overall R$^2$ improving from 0.49 to 0.75 (NO$_2$) and from 0.50 to 0.80 (PM$_{2.5}$). In terms of predictively ability, reductions of 2.4$\mu$g/m$^3$ (24.2\%) and 2.2$\mu$g/m$^3$ (19.5\%) in RMSE and PwRMSE respectively, are observed in NO$_2$ with reductions of 1.8$\mu$g/m$^3$ (50.0\%) and 1.5$\mu$g/m$^3$ (44.1\%) in RMSE and PwRMSE respectively for PM$_{2.5}$. \\

\noindent Furthermore, allowing spatio-temporal variation (Model (iv)) in the both the intercept and the coefficient associated with the CTM results in further improvements, particularly in PM$_{2.5}$, in all metrics (for both within- and out-of-sample model fit) compared to a model that only allows spatial variation in the coefficients (Model ii)). In terms of model fit, we see overall R$^2$ improving from  0.75 to 0.77 (NO$_2$) and from 0.80 to 0.93 (PM$_{2.5}$), with the out-of-sample RMSE and PwRMSE by 0.2$\mu$g/m$^3$ (2.7\%) and 0.3$\mu$g/m$^3$ (3.2\%) respectively for NO$_2$ and decreasing by a further 0.8$\mu$g/m$^3$ (30.8\%) and 0.7$\mu$g/m$^3$ (26.9\%) for PM$_{2.5}$. \\ 

\begin{table}[t]
\footnotesize
\centering
	\caption{Summary of results from fitting the three candidate models. Results are presented for both in-sample model fit and out-of-sample predictive ability and are the median values from 25 training-validation set combinations. For both within- and out-of- sample model fit, R$^2$, root mean squared error (RMSE) and population weighted root mean squared error (PwRMSE) are given.}\label{tab::evaluation}
	\begin{tabular}{cccc ccc cc}
	\hline \\[-8pt]
	 & & \multicolumn{3}{c}{\textbf{Within-sample}}  & \multicolumn{3}{c}{\textbf{Out-of-sample}}  \\[1pt] 
	 \cmidrule(lr){3-5}
	 \cmidrule(lr){6-8}
	\textbf{Model} & & \textbf{R$^2$} & \textbf{RMSE} & \textbf{PwRMSE} & \textbf{R$^2$} & \textbf{RMSE} & \textbf{PwRMSE} \\[3pt] 
	\hline \\[-5pt]
		\multicolumn{2}{l}{NO$_{2}$} &&&&&\\
	  \multicolumn{2}{c}{$(i)$} & 0.49 & 9.9 & 11.4 & 0.49 & 9.9 & 11.3 \\
	  \multicolumn{2}{c}{$(iii)$} & 0.75 & 7.0 & 8.9 & 0.70 & 7.5 & 9.4 \\ 
	  \multicolumn{2}{c}{$(iv)$} & 0.77 & 6.7 & 8.6 & 0.72 & 7.3 & 9.1 \\[5pt]
		\multicolumn{2}{l}{PM$_{2.5}$} &&&&&\\
	  \multicolumn{2}{c}{$(i)$} & 0.50 & 3.5 & 3.4 & 0.49 & 3.6 & 3.4 \\
	  \multicolumn{2}{c}{$(iii)$} & 0.80 & 2.3 & 2.4 & 0.74 & 2.6 & 2.6 \\
	  \multicolumn{2}{c}{$(iv)$} & 0.93 & 1.3 & 1.5 & 0.87 & 1.8 & 1.9 \\[5pt]
	\hline
	\end{tabular}
\end{table}

\noindent Full posterior distributions from the final model (Model (iv)) were produced for each cell on a high-resolution grid (comprising of approximately 3.7 million, 1km $\times$ 1km cells) covering Western Europe. These can be used to produce high-resolution maps of concentrations, maps of exceedances, changes over time and, by aligning with high--resolution estimates of populations to create estimates of trends in country-level population-weighted annual average concentrations of NO$_2$ and PM$_{2.5}$. \\

\begin{figure}[H]
	\centering
	\begin{subfigure}{0.49\linewidth}  \centering
 		\includegraphics[width = \linewidth, clip = TRUE, trim = {4cm 2cm 4cm 2cm}]{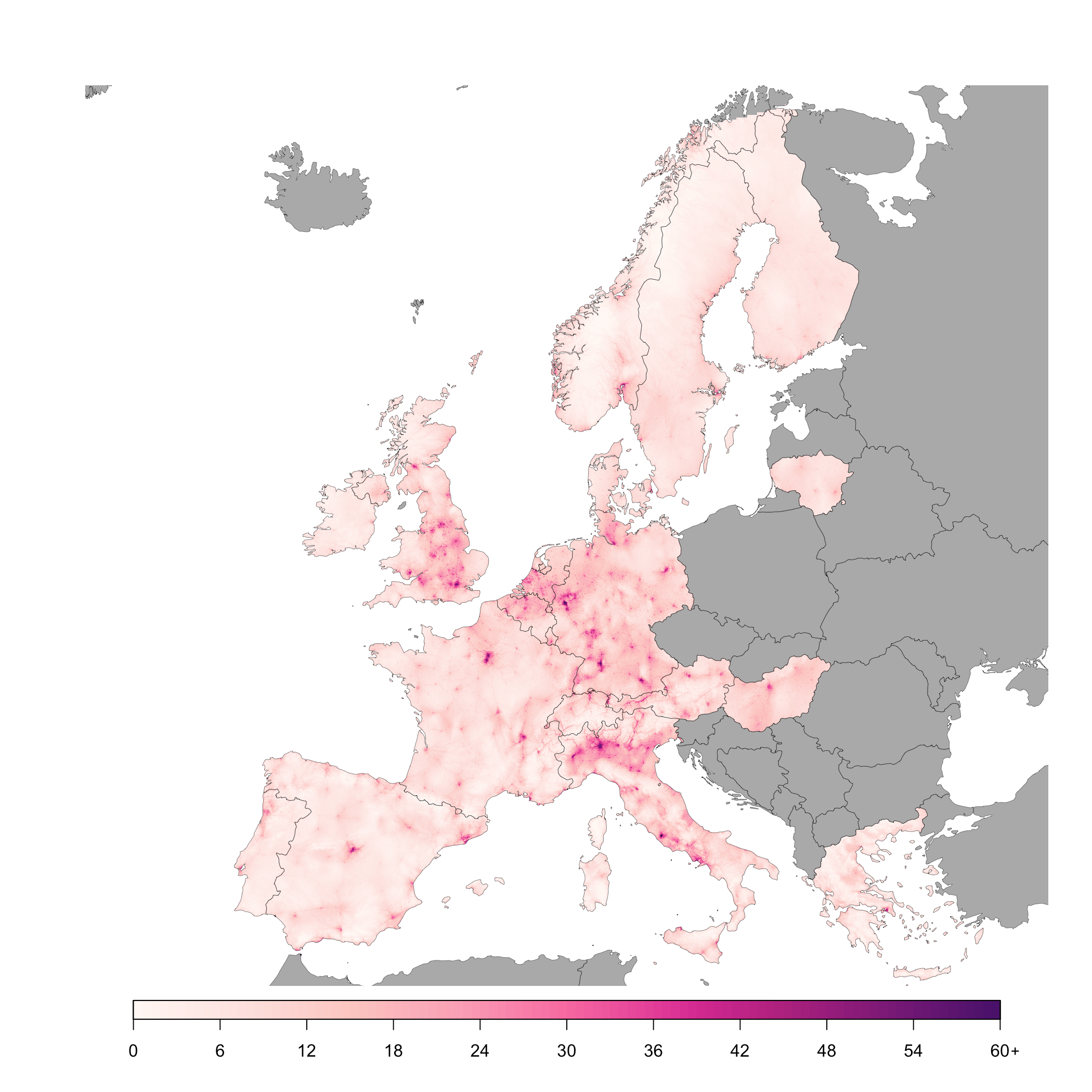}
 		\caption{Median NO$_2$} 
 	\end{subfigure}
	\begin{subfigure}{0.49\linewidth}  \centering
 		\includegraphics[width = \linewidth, clip = TRUE, trim = {4cm 2cm 4cm 2cm}]{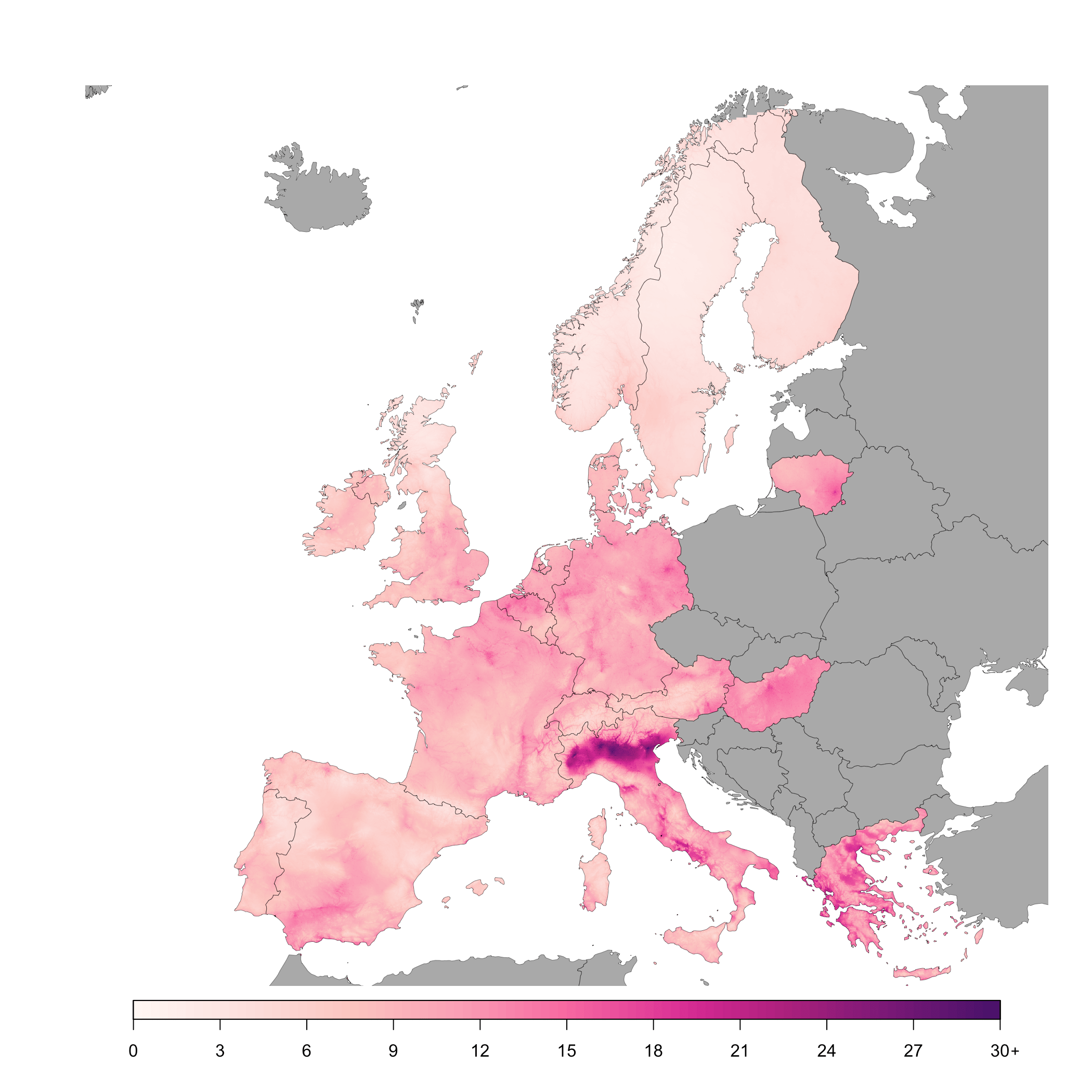}
 		\caption{Median PM$_{2.5}$} 
 	\end{subfigure}
	\caption{Median annual average concentrations of NO$_2$ and PM$_{2.5}$ (in $\mu$g/m$^{3}$) in 2016, by grid-cell (1km $\times$ 1km resolution)}
 		\label{fig:mapmed}
\end{figure}

\begin{figure}[H]
	\centering
 		\begin{subfigure}{0.49\linewidth}  \centering
 		\includegraphics[width = \linewidth, clip = TRUE, trim = {4cm 2cm 4cm 2cm}]{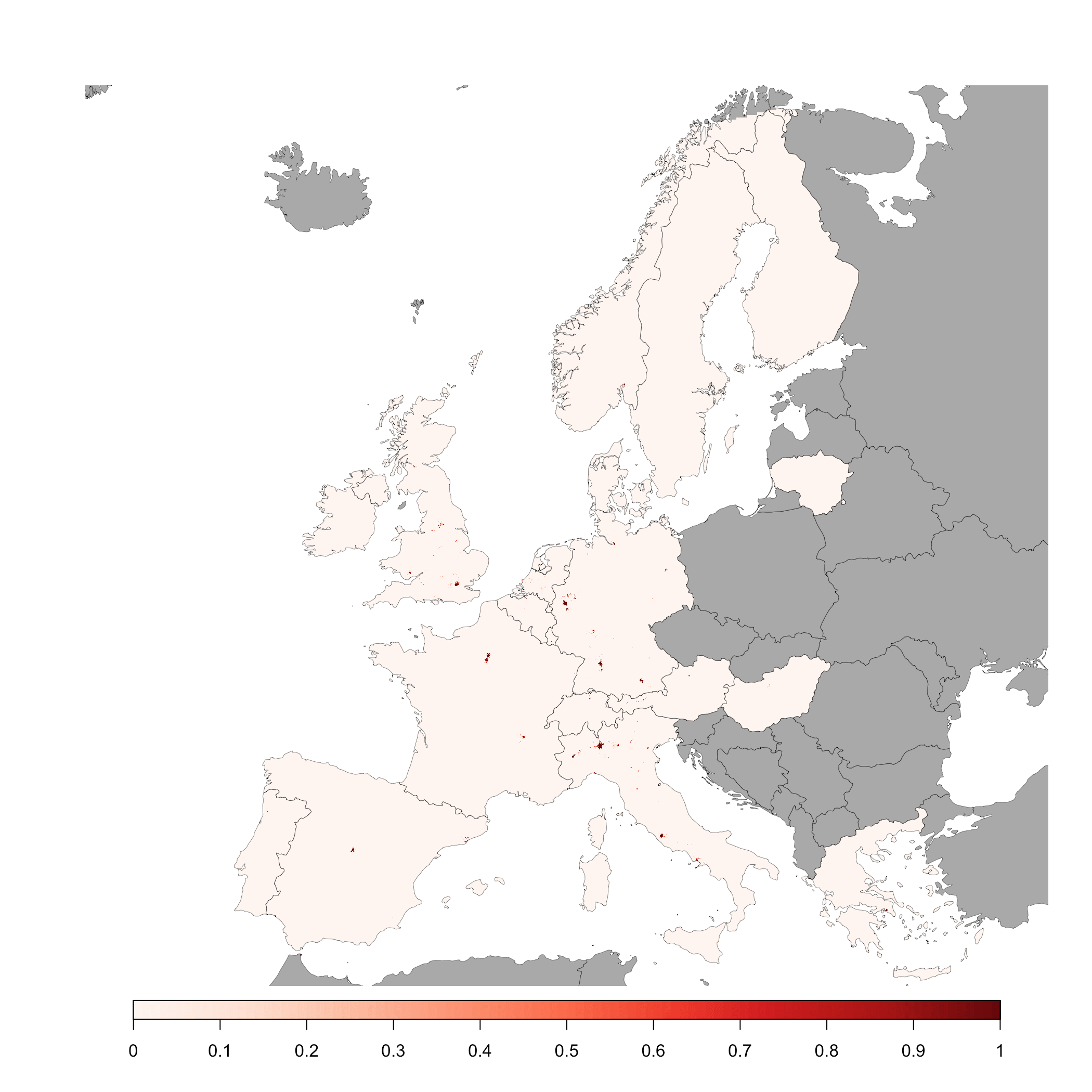}
 		\caption{Probability NO$_2$ exceeds 40$\mu$g/m$^{3}$} 
 	\end{subfigure}
	\begin{subfigure}{0.49\linewidth}  \centering
 		\includegraphics[width = \linewidth, clip = TRUE, trim = {4cm 2cm 4cm 2cm}]{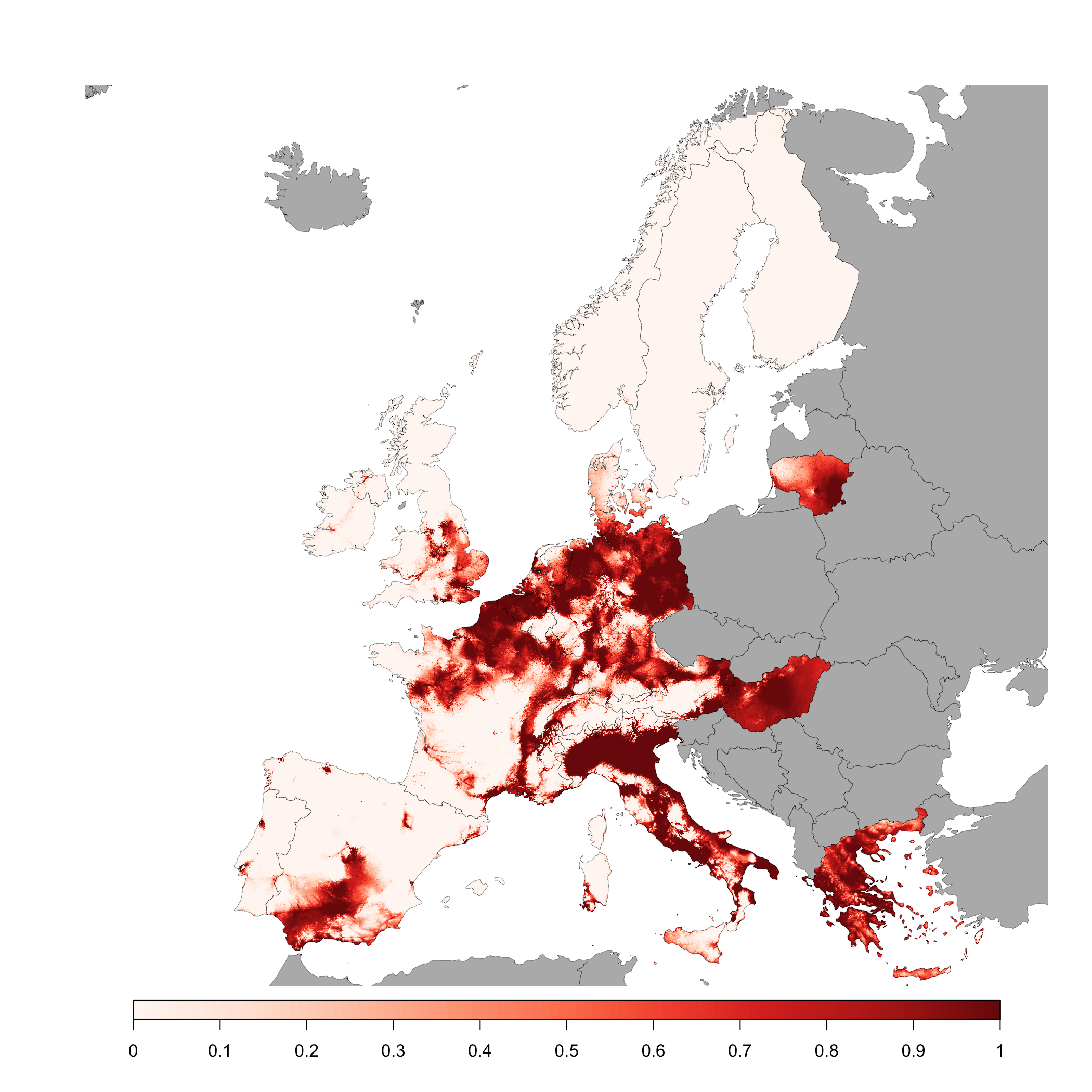}
 		\caption{Probability PM$_{2.5}$ exceeds 10$\mu$g/m$^{3}$} 
 	\end{subfigure}
	\caption{Probability that the annual average concentration of NO$_2$ and PM$_{2.5}$ (in $\mu$g/m$^{3}$) in 2016 exceed the WHO AQGs, by grid-cell (1km $\times$ 1km resolution).}
	\label{fig::mapexceed}
\end{figure}

\noindent Figure \ref{fig:mapmed} show maps of estimated annual average concentrations (at a 1km $\times$ 1km resolution) of (a) NO$_2$ and (b) PM$_{2.5}$ in 2016 (medians of the posterior distribution in each grid-cell) respectively. Probability of exceedances can be seen in Figure \ref{fig::mapexceed} which shows the estimated probability that the annual average (a) NO$_2$ and (b) PM$_{2.5}$ in 2016 exceed that of the WHO AQGs of 40$\mu$g/m$^{3}$ and 10$\mu$g/m$^{3}$ respectively. Higher levels of both pollutants are seen in urban areas, however, overall the level of spatial variation are different. NO$_2$ is mainly derived from transport and other combustion sources and, as can be seen in Figure \ref{fig:mapmed}(a), the highest concentrations are seen in areas with high levels of traffic and/or high population density such as the larger metropolitan areas in Europe including London, Paris and Cologne. This is also seen in Figure \ref{fig::mapexceed}(a), where the highest probabilities are confined to large cities where emissions from road transport and other combustion sources would be expected to be highest. In contrast, PM$_{2.5}$ is derived from a wide range of sources, including energy production, industry, transport, agriculture, dust and forest fires and it is known that particles can travel in the atmosphere over long distances. This can be seen in Figure \ref{fig:mapmed}(b)  where concentrations are much more widespread with the highest observed in Northern Italy, a region with a high population and industrial activity. Figure \ref{fig::mapexceed}(b) also shows that concentrations are more widespread with high probabilities of exceeding the WHO AQGs occurring in both urban and rural areas. \\ 

\begin{figure}[H]
	\centering
 	\includegraphics[width = 0.85\linewidth]{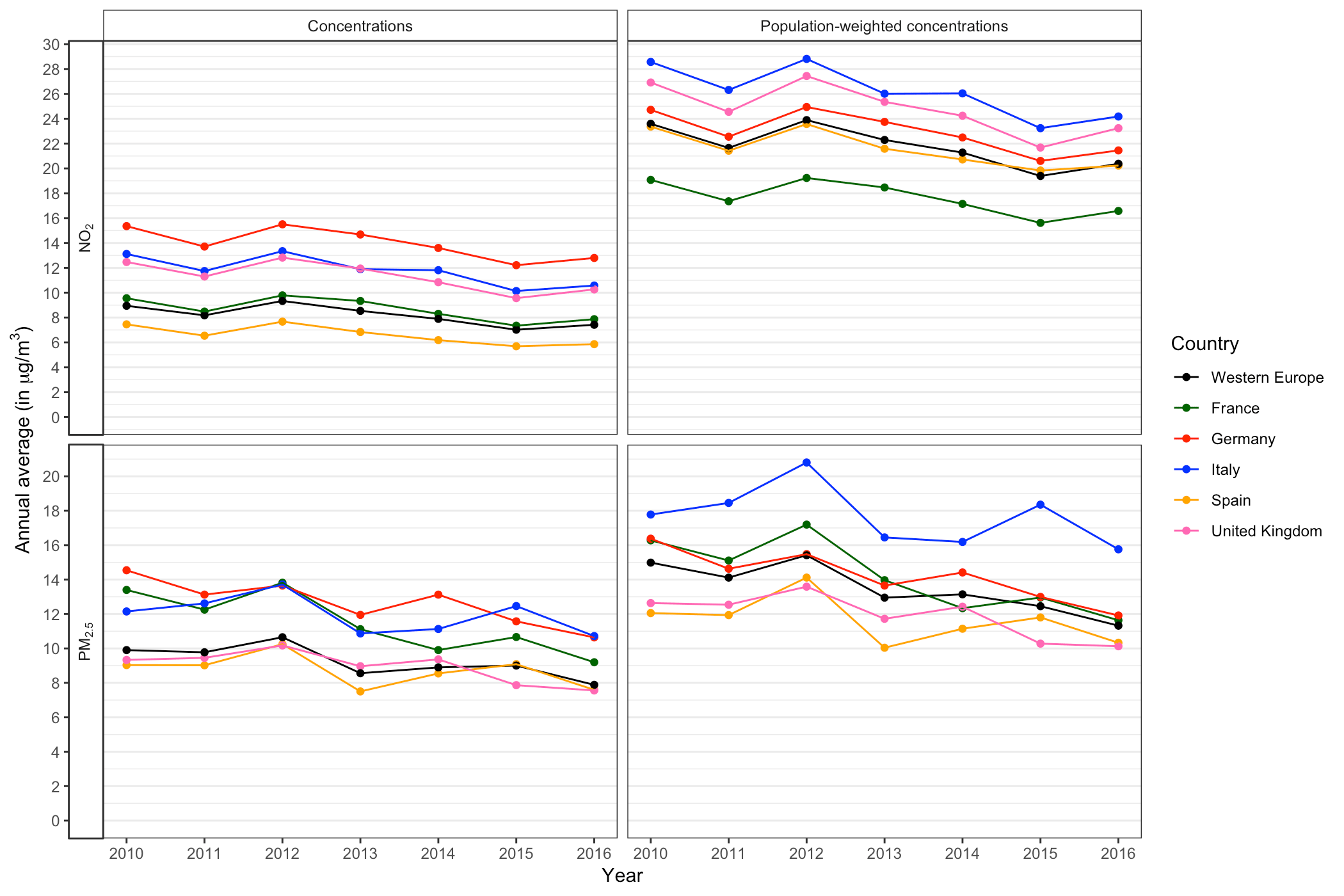}
	\caption{Trends in concentrations and population-weighted concentrations  for NO$_2$ and PM$_{2.5}$ between 2010 and 2016 for the five most populated countries, together with overall trend in Western Europe.}\label{fig::timeseries}
\end{figure}

\noindent Figure \ref{fig::timeseries} shows trends in annual average concentrations and population-weighted annual average concentrations between 2010 and 2016 for the five most populated countries in Europe along with the overall trend in Western Europe (corresponding trends for all countries together with prediction intervals can be found in Appendix Table \ref{tab:annavg}). Overall, estimated concentrations of both NO$_2$ and PM$_{2.5}$ decreased between 2010 to 2016, with the rate of decline for NO$_2$ appears relatively consistent for all countries with much more variability observed in the trends of PM$_{2.5}$. It can also be seen that population-weighted annual average concentrations are higher than the unweighted annual average concentrations indicating that areas with high annual average concentrations are typically co-located with areas with high populations. \\

\begin{figure}[H]
	\centering
	\begin{subfigure}{0.49\linewidth}  \centering
 		\includegraphics[width = \linewidth, clip = TRUE, trim = {4cm 2cm 4cm 2cm}]{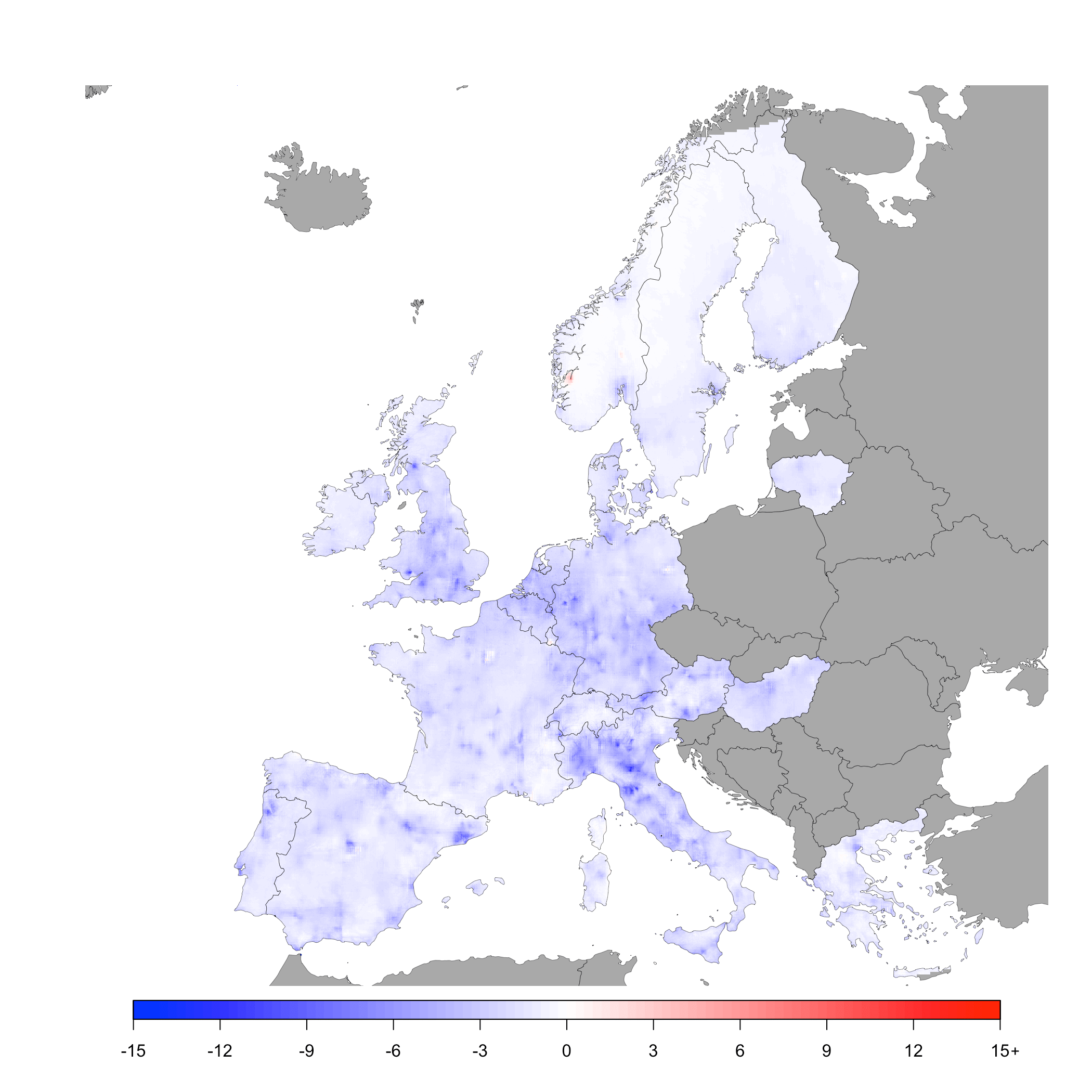}
 		\caption{NO$_2$} 
 	\end{subfigure}
	\begin{subfigure}{0.49\linewidth}  \centering
 		\includegraphics[width = \linewidth, clip = TRUE, trim = {4cm 2cm 4cm 2cm}]{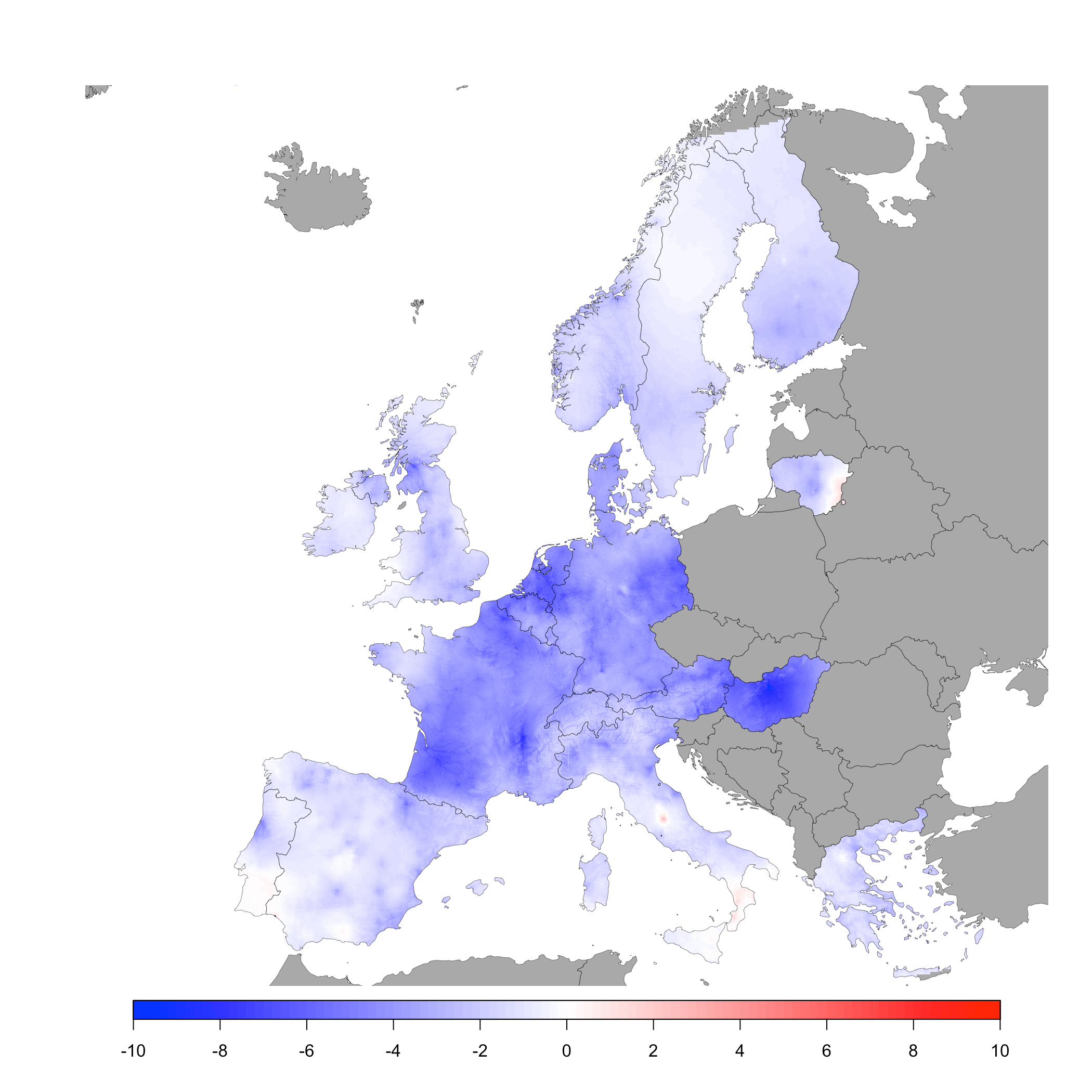}
 		\caption{PM$_{2.5}$} 
 	\end{subfigure}
	\caption{Median difference in annual average concentrations of NO$_2$ and PM$_{2.5}$ (in $\mu$g/m$^{3}$) between 2010 and 2016, by grid-cell (1km $\times$ 1km resolution).}\label{fig::mapchange}
\end{figure}

\noindent Figure \ref{fig::mapchange} shows the differences in annual average concentrations of (a) NO$_2$ and (b) PM$_{2.5}$ between 2010 and 2016, for each grid-cell (1km $\times$ 1km resolution), the median of 1000 samples is presented. Each sample represents the difference between the (samples from) the joint posterior distributions for 2016 and 2010. Again, we see that  concentrations of NO$_2$ and PM$_{2.5}$ decreased between 2010 to 2016, across Western Europe. Across Western Europe, the median decrease in NO$_2$ and PM$_{2.5}$ was 1.4 and 2.0 $\mu$g/m$^{3}$ respectively, with these decreases increasing to 2.9 and 3.3 $\mu$g/m$^{3}$ respectively when weighting by population, indicating that air quality is improving in areas with higher population (See Appendix Table \ref{tab::summdiff}). The largest country--level decreases in PM$_{2.5}$ are seen in Hungary (6.0$\mu$g/m$^{3}$), the Netherlands (5.3$\mu$g/m$^{3}$) and France (4.1$\mu$g/m$^{3}$) for PM$_{2.5}$ and largest country--level decreases in PM$_{2.5}$ seen in the Netherlands (2.9$\mu$g/m$^{3}$), Belgium (2.9$\mu$g/m$^{3}$) and Germany (2.5$\mu$g/m$^{3}$), again with these decreases increasing when population--weighted (See Appendix Table \ref{tab::summdiff}). The smallest decreases in the annual average concentrations and population-weighted concentrations of NO$_2$ and PM$_{2.5}$ seen in Norway, Sweden and Finland, however concentrations are generally lower in these countries and it reflects the difficulty of reducing air pollution at low levels.\\

\begin{figure}[H]
	\centering
	\begin{subfigure}{0.49\linewidth}
		\centering	
		\includegraphics[width = 0.8\linewidth]{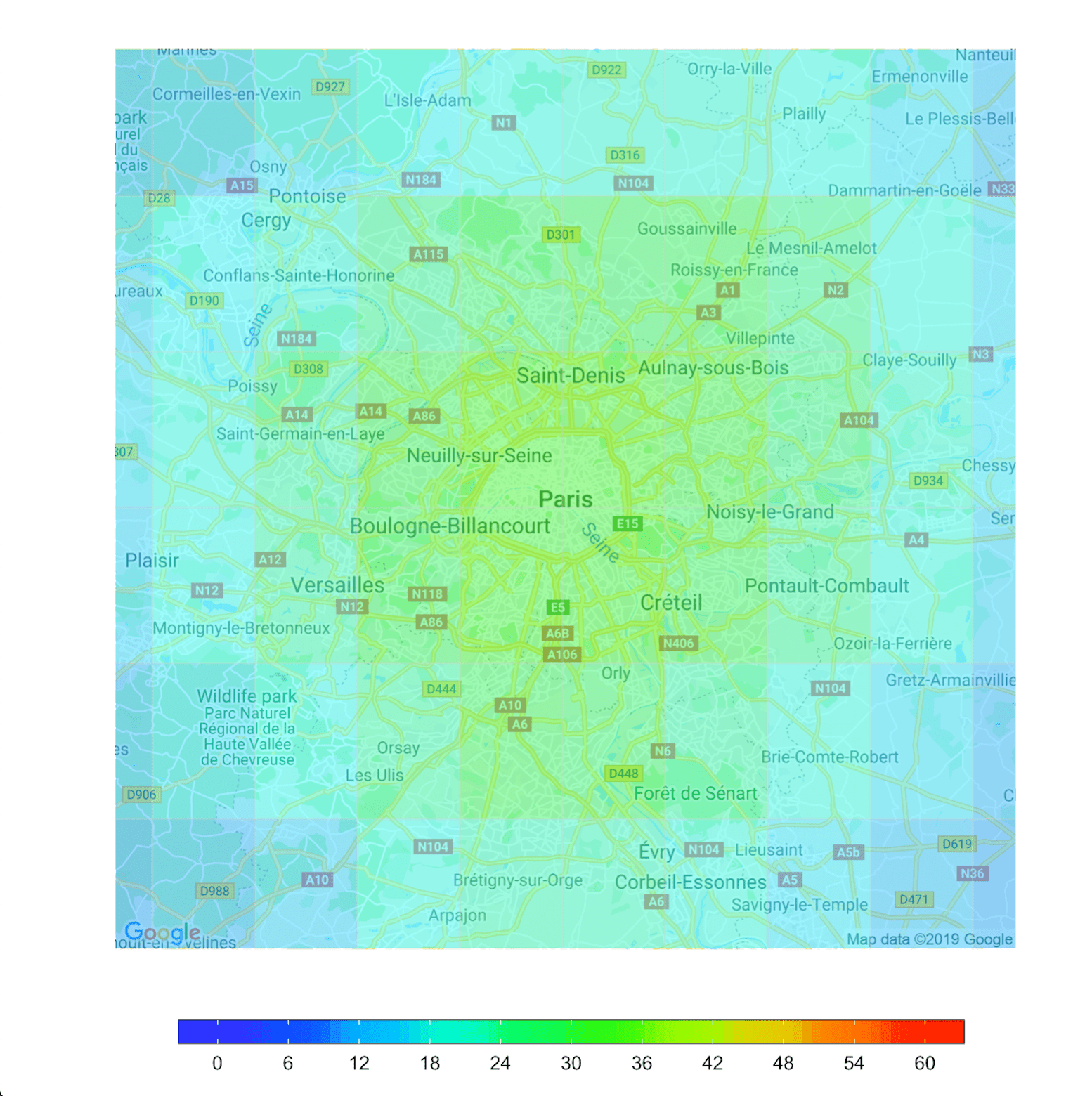}
		\caption{Estimated annual average concentrations of NO$_{2}$ from the MACC-II CTM (10km $\times$ 10km resolution).}
	\end{subfigure}
	\begin{subfigure}{0.49\linewidth}
		\centering	
		\includegraphics[width = 0.8\linewidth]{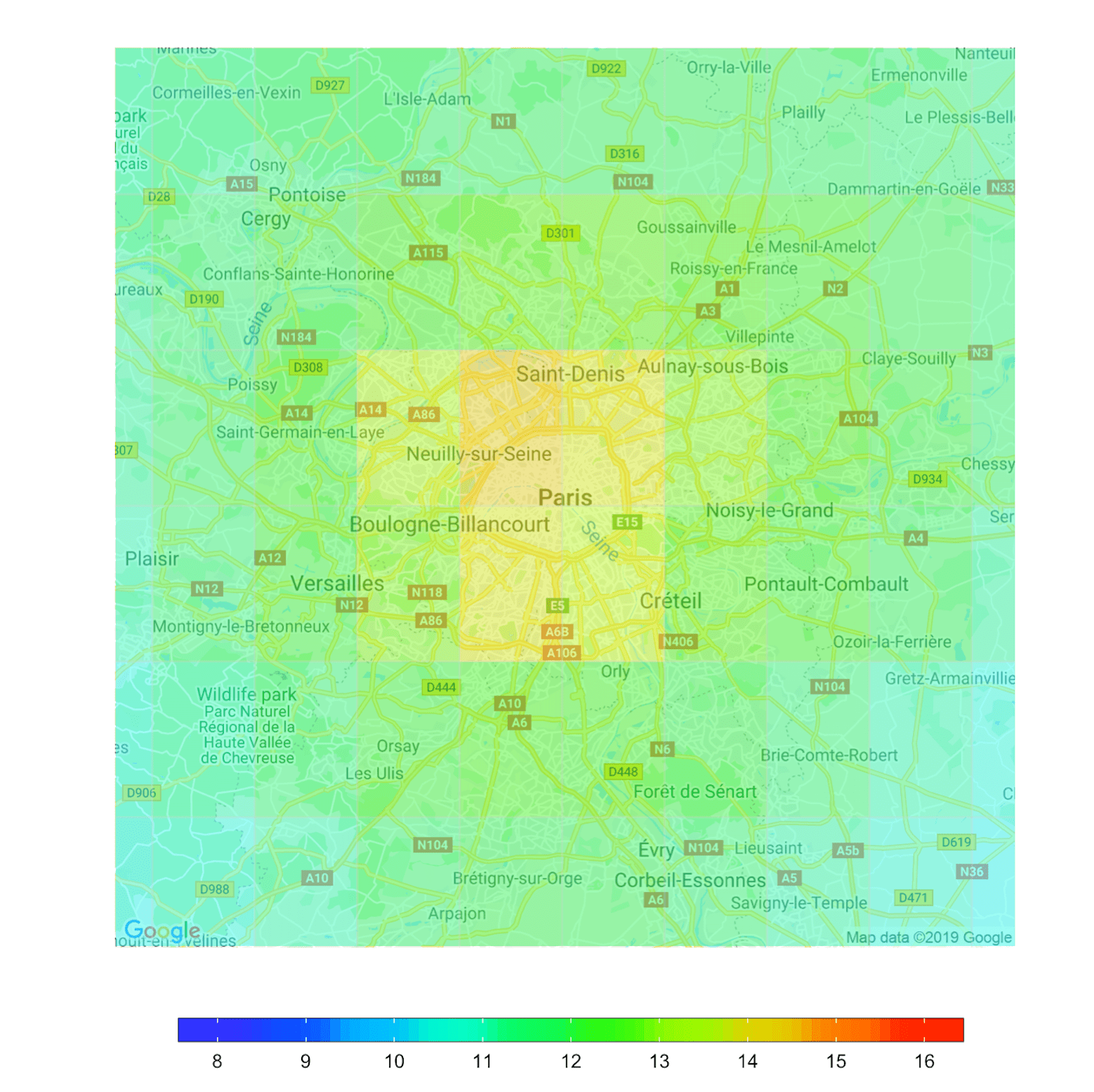}
		\caption{Estimated annual average concentrations of PM$_{2.5}$ from the MACC-II CTM (10km $\times$ 10km resolution).}
	\end{subfigure}
	\begin{subfigure}{0.49\linewidth}
		\centering	
		\includegraphics[width = 0.8\linewidth]{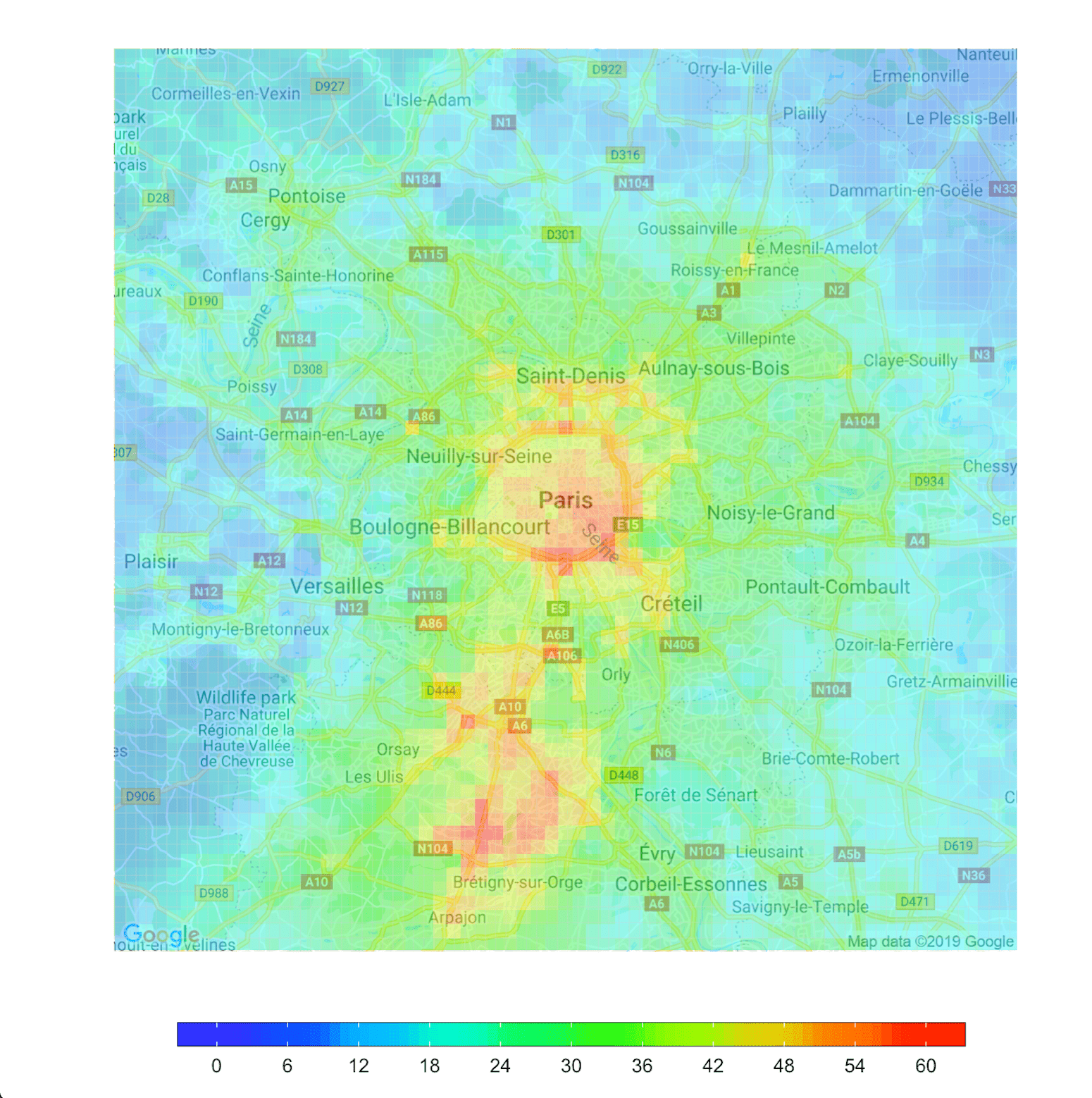}
		\caption{Median annual average concentrations of NO$_2$ (1km $\times$ 1km resolution).}
	\end{subfigure}
	\begin{subfigure}{0.49\linewidth}
		\centering	
		\includegraphics[width = 0.8\linewidth]{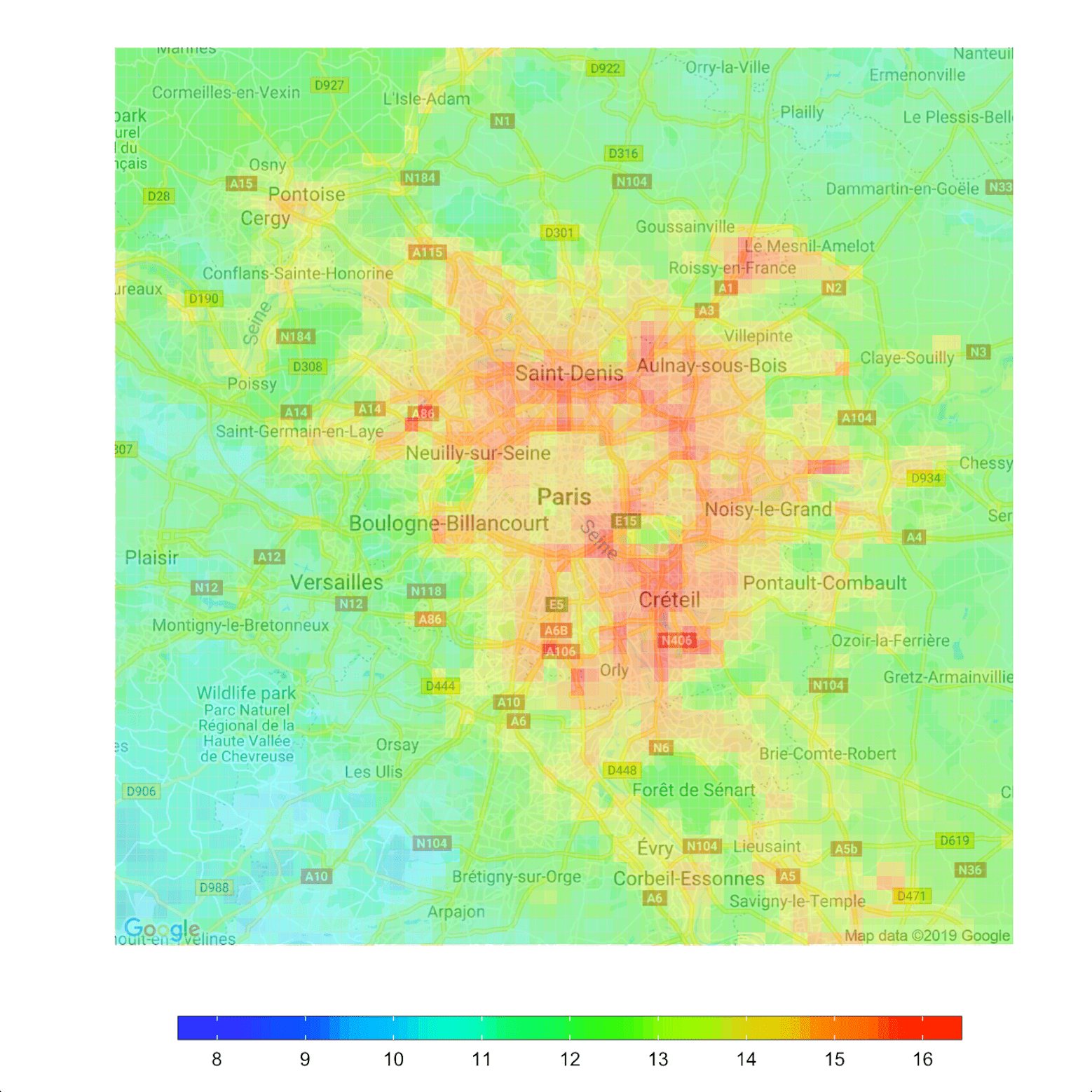}
		\caption{Median annual average concentrations of PM$_{2.5}$ (1km $\times$ 1km resolution).}
	\end{subfigure}
	\begin{subfigure}{0.49\linewidth}
		\centering	
		\includegraphics[width = 0.8\linewidth]{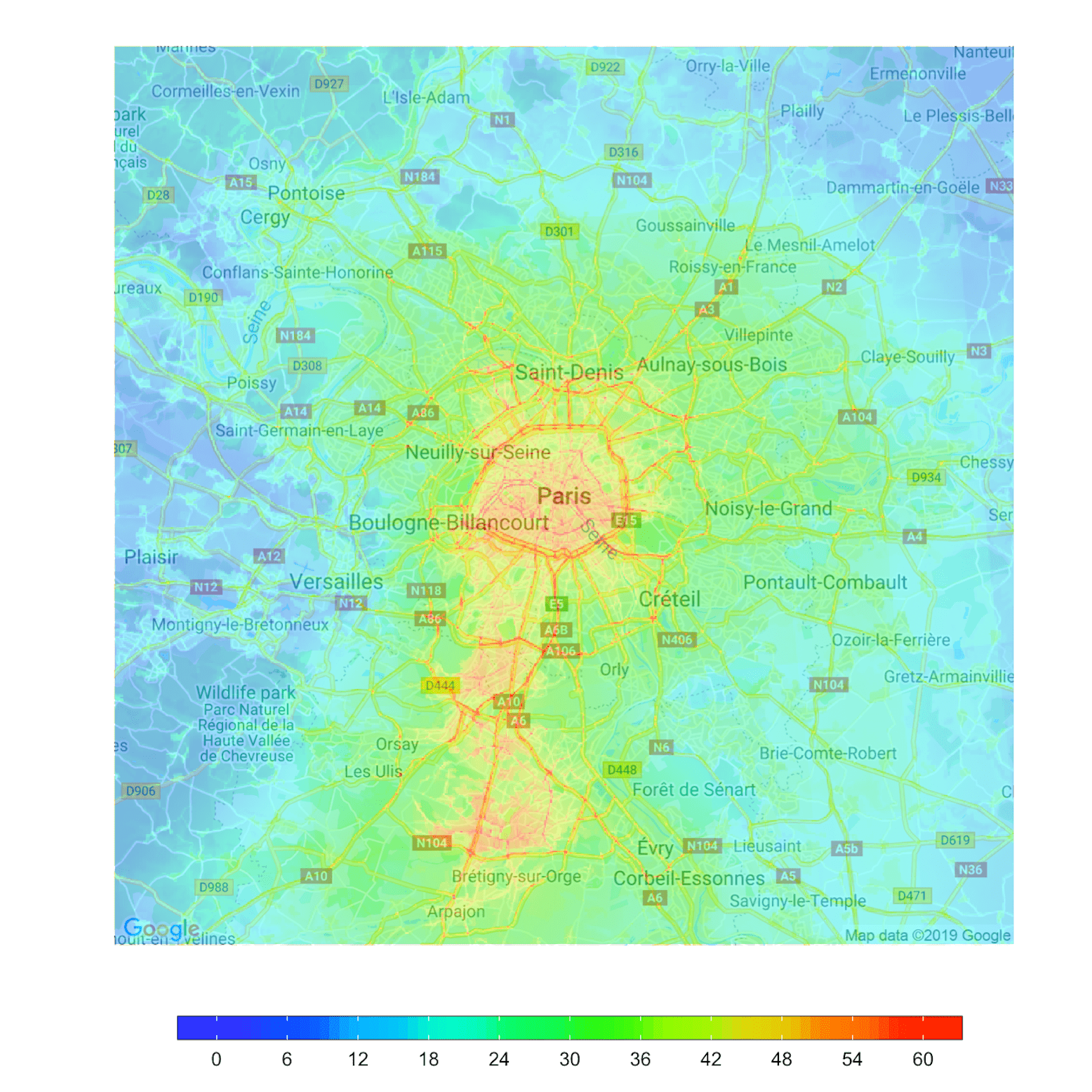}
		\caption{Median annual average concentrations of NO$_2$ (100m $\times$ 100m resolution).}
	\end{subfigure}
	\begin{subfigure}{0.49\linewidth}
		\centering	
		\includegraphics[width = 0.8\linewidth]{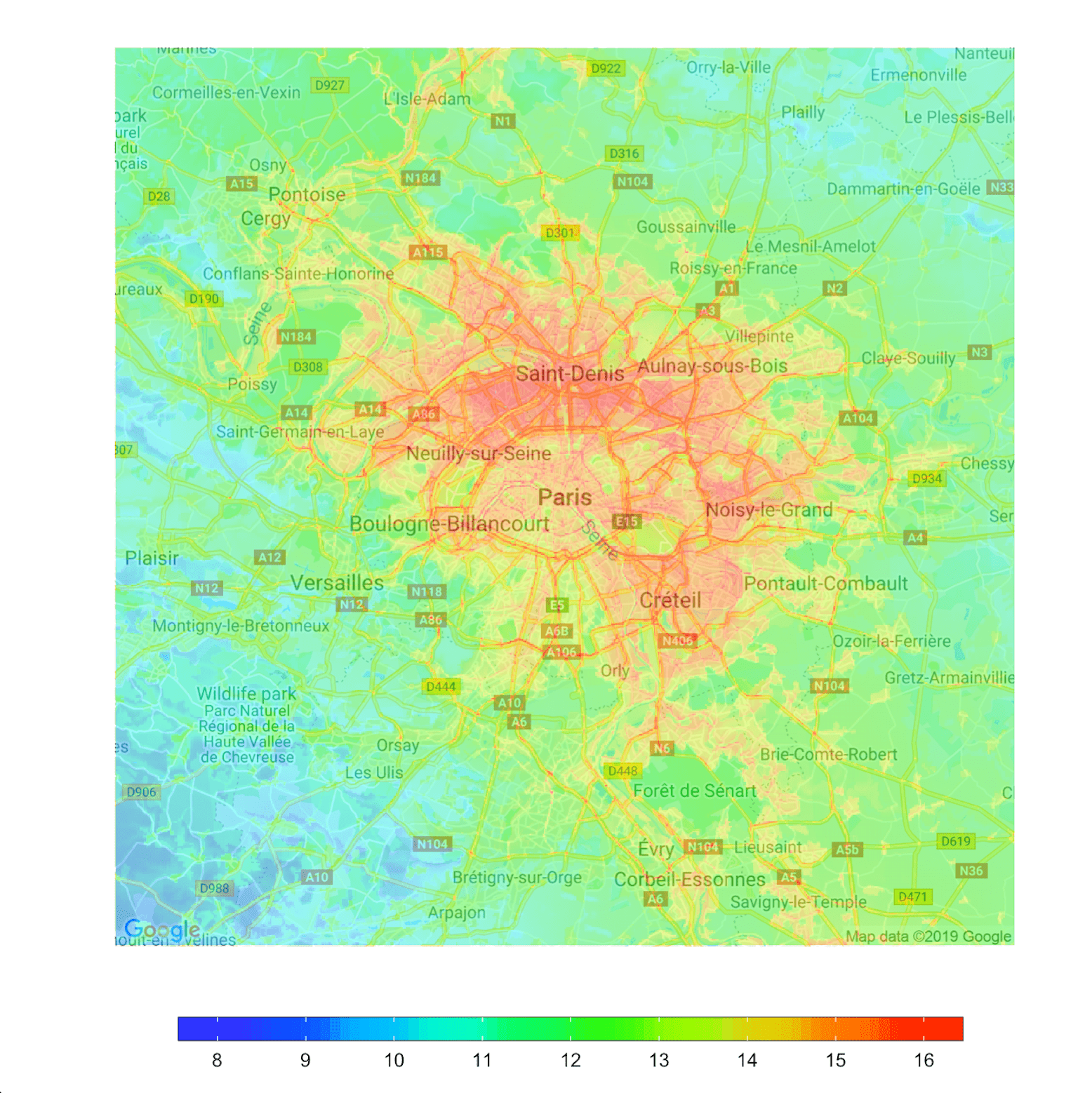}
		\caption{Median annual average concentrations of PM$_{2.5}$ (100m $\times$ 100m resolution).}
	\end{subfigure}
	\caption{Estimated annual averages of NO$_2$ and PM$_{2.5}$ (in $\mu$g/m$^{3}$) in 2016 for Paris by grid cell.} \label{fig::granular}
\end{figure}

\noindent The effect of using higher resolution land use information (100m $\times$ 100m) in the model can be seen in Figure \ref{fig::granular} (e) and (f). This shows the ability of the model to capture within city variation and produce high-resolution estimates of concentrations of pollutants over space and time and to capture within-city variation, compared to 1km $\times$ 1km (Figure \ref{fig::granular}(c) and Figure \ref{fig::granular}(d)). Using the model with land use information at either 100m $\times$ 100m or 1km $\times$ 1km resolution shows a clear increase in granularity from the MACC-II Ensemble CTM (Figure \ref{fig::granular}(a) and Figure \ref{fig::granular}(b)) and provides substantial additional information on the spatial distribution of NO$_2$ and PM$_{2.5}$ that can be used for policy support, assigned to residential addressed for use in epidemiological analyses and matching to populations to estimate the burden of air pollution on human health. 

%%%%%%%%%%%%%%%%%%%%%%%%%%%%%%%%%%%%%%%%%%%%%%%%%%%%%%%%%%%%%%%%%%
%%%%%%%%%%%%%%%%%%%%%%%%%%%%%%%%%%%%%%%%%%%%%%%%%%%%%%%%%%%%%%%%%%
%%%%%%%%%%%%%%%%%%%%%%%%%%%%%%%%%%%%%%%%%%%%%%%%%%%%%%%%%%%%%%%%%%

\section{Discussion}\label{sec::discuss} 

%%%%%%%%%%%%%%%%%%%%%%%%%%%%%%%%%%%%%%%%%%%%%%%%%%%%%%%%%%%%%%%%%%
%%%%%%%%%%%%%%%%%%%%%%%%%%%%%%%%%%%%%%%%%%%%%%%%%%%%%%%%%%%%%%%%%%
%%%%%%%%%%%%%%%%%%%%%%%%%%%%%%%%%%%%%%%%%%%%%%%%%%%%%%%%%%%%%%%%%%

\noindent The ability to produce comprehensive, accurate, high-resolution, estimates of concentrations of pollutants is essential for policy support, epidemiological analyses and in estimating the burden of air pollution on human health.  Traditionally, the primary source of information on air pollution has come from ground monitoring networks but these may not always be able to provide sufficient spatial coverage. In such cases, information from ground measurements can be combined with other sources that provide comprehensive estimates but these cannot capture micro-scale features that may be reflected in the ground measurements and may be important in assessing individuals exposure to harmful levels of air pollution. \\

\noindent  When combining data from different sources, it is important to acknowledge possible differences in the spatial  resolutions (or changes in support) between data sources. Here, we use a model that calibrates estimates from a chemical transport model along with land use information, road network information and topography with ground measurements (where available). Set within a Bayesian hierarchical framework, calibration equations link the data sources with ground measurements, allowing these relationships to vary continuously over space and time. In the example presented, we use the model to produce posterior distributions of annual average concentrations of NO$_2$ and PM$_{2.5}$ for each of 3.7 million grid-cells at a 1km$\times$ 1km resolution between 2010 and 2016 and demonstrated the ability to produce estimates at an even higher resolution (100m $\times$ 100m). We specifically address the computational issues that arise when fitting spatial and temporally varying coefficient models for larger scale problems by performing approximate Bayesian inference based on integrated nested Laplace approximations. The result is the ability to produce a comprehensive set of high-resolution estimates, together with associated measures of uncertainty, for the entire study region. \\

\noindent   High concentrations of  NO$_2$ were observed in urban areas, with the highest levels (those above the WHO AQG of an annual average of 40 $\mu$g/m$^{3}$) being seen around the major cities across Europe. Spatial patterns in estimated concentrations of PM$_{2.5}$ were much smoother (than for NO$_2$) with high levels seen across large parts of  Europe, including many rural areas. Overall, estimated concentrations of both NO$_2$ and PM$_{2.5}$  decreased during the period 2010 to 2016 with the largest decreases being seen for PM$_{2.5}$. When examining trends in population-weighted concentrations,  for NO$_2$ the rate of decline appears relatively consistent for all countries with much more variability observed in the trends of PM$_{2.5}$. \\

\noindent In the models used here, there is an assumption the covariates are error free, an assumption that in practice may be untenable. Estimates of pollutant concentrations from  numerical simulations of atmospheric models are the result of modelling and, as such, will be subject to uncertainties and biases arising from errors in inputs and possible model misspecification. An alternative approach to statistical calibration is Bayesian melding \citep{poole2000inference}. In melding, both the measurements and the estimates are assumed to arise from an underlying latent process that represents the true level of pollution. This latent process itself is unobservable but measurements can be taken at locations in space and time. For example, the underlying latent process would represent the true level of PM$_{2.5}$ and  drives the measurements from ground monitors and the estimates from the CTM, all of which will inform the posterior distribution of the underlying latent process. At present, implementing Bayesian melding can be very computationally demanding, particularly using MCMC, and future research is required to develop computationally efficient methods for implementation that would allow its use in practical settings, such as the one considered here. This will allow information from a wide variety of sources, at different spatial aggregations each with different uncertainties and error structures, to be integrated within a flexible and coherent framework.

% \noindent There may be issues when using regular lattice over large spatial domains, due to the curvature of the earth. For example, a 0.1$^{\degree}$ lattice will produce cells of approximately 11km$\times$11km at the equator but will decrease the closer to the poles. Hence the relationship between correlation and distance will not be constant over space and the underlying notion of distance will differ according to location. An additional difficulty presents itself when modelling air pollution on a global setting as the density of monitoring networks varies substantially across the globe \citep{Shaddick2017DIMAQ}. The ability to detect small range spatial variation will be restricted in areas where monitoring is sparse and therefore the tendency will be for the range parameter to be large representing larger scale (as points will be generally further apart). \\

%%%%%%%%%%%%%%%%%%%%%%%%%%%%%%%%%%%%%%%%%%%%%%%%%%%%%%%%%%%%%%%%%%
%%%%%%%%%%%%%%%%%%%%%%%%%%%%%%%%%%%%%%%%%%%%%%%%%%%%%%%%%%%%%%%%%%
%%%%%%%%%%%%%%%%%%%%%%%%%%%%%%%%%%%%%%%%%%%%%%%%%%%%%%%%%%%%%%%%%%

\section*{Acknowledgments}

%%%%%%%%%%%%%%%%%%%%%%%%%%%%%%%%%%%%%%%%%%%%%%%%%%%%%%%%%%%%%%%%%%
%%%%%%%%%%%%%%%%%%%%%%%%%%%%%%%%%%%%%%%%%%%%%%%%%%%%%%%%%%%%%%%%%%
%%%%%%%%%%%%%%%%%%%%%%%%%%%%%%%%%%%%%%%%%%%%%%%%%%%%%%%%%%%%%%%%%%

Matthew L. Thomas is supported by a scholarship from the Engineering and Physical Sciences Research Council Centre for Doctoral Training in Statistical Applied Mathematics at Bath, under project EP/L015684/1. The authors would like to acknowledge the use of the University of Exeter High-Performance Computing (HPC) facility in carrying out this work.

%%%%%%%%%%%%%%%%%%%%%%%%%%%%%%%%%%%%%%%%%%%%%%%%%%%%%%%%%%%%%%%%%%
%%%%%%%%%%%%%%%%%%%%%%%%%%%%%%%%%%%%%%%%%%%%%%%%%%%%%%%%%%%%%%%%%%
%%%%%%%%%%%%%%%%%%%%%%%%%%%%%%%%%%%%%%%%%%%%%%%%%%%%%%%%%%%%%%%%%%

\bibliographystyle{mattnat}
\bibliography{Refs}

%%%%%%%%%%%%%%%%%%%%%%%%%%%%%%%%%%%%%%%%%%%%%%%%%%%%%%%%%%%%%%%%%%
%%%%%%%%%%%%%%%%%%%%%%%%%%%%%%%%%%%%%%%%%%%%%%%%%%%%%%%%%%%%%%%%%%
%%%%%%%%%%%%%%%%%%%%%%%%%%%%%%%%%%%%%%%%%%%%%%%%%%%%%%%%%%%%%%%%%%

%\begin{figure}[H]
%	\centering
%	\includegraphics[scale = 0.8, ]{Figures/SimStudyResults.pdf}
%	\caption{Results of simulation study. True values of parameters are denoted by red dotted lines with estimates shown by dots. Dots signify the median of the estimates from using multiple simulated datasets with the vertical lines showing the range of the estimates. Values are given for three different models: (i) intercept and association with CTM allowed to vary over space; (ii) intercept allowed to vary over space but association with CTM fixed; (iii) both are fixed (over space); see text for details.
%	\label{fig::simres}}
%\end{figure}

%%%%%%%%%%%%%%%%%%%%%%%%%%%%%%%%%%%%%%%%%%%%%%%%%%%%%%%%%%%%%%%%%%
%%%%%%%%%%%%%%%%%%%%%%%%%%%%%%%%%%%%%%%%%%%%%%%%%%%%%%%%%%%%%%%%%%
%%%%%%%%%%%%%%%%%%%%%%%%%%%%%%%%%%%%%%%%%%%%%%%%%%%%%%%%%%%%%%%%%%

\begin{appendix}

\section{Appendix}
\linespread{1}

%%%%%%%%%%%%%%%%%%%%%%%%%%%%%%%%%%%%%%%%%%%%%%%%%%%%%%%%%%
%%%% Resetting figure and table count for the appendix %%%
%%%%%%%%%%%%%%%%%%%%%%%%%%%%%%%%%%%%%%%%%%%%%%%%%%%%%%%%%%
\setcounter{figure}{0} \renewcommand{\thefigure}{A.\arabic{figure}}
\setcounter{table}{0} \renewcommand{\thetable}{A.\arabic{table}}

%%%%%%%%%%%%%%%%%%%%%%%%%%%%%%%%%%%
%%%% Table of model fit by year %%%
%%%%%%%%%%%%%%%%%%%%%%%%%%%%%%%%%%%
\begin{table}[H]
\small
\centering
	\caption{Summary of results from fitting the three candidate models by year. Results are presented for both in-sample model fit and out-of-sample predictive ability and are the median values from 25 training-validation set combinations. For both within- and out-of- sample model fit, R$^2$, root mean squared error (RMSE) and population weighted root mean squared error (PwRMSE) are given.}\label{tab::evaluationbyyear}
	\begin{tabular}{cccc ccc cc}
	\hline \\[-8pt]
	 & & & \multicolumn{3}{c}{\textbf{Model Fit}}  & \multicolumn{3}{c}{\textbf{Predictive Accuracy}}  \\[1pt] 
	 \cmidrule(lr){4-6}
	 \cmidrule(lr){7-9}
	\textbf{Model} & & \textbf{Year} & \textbf{R$^2$} & \textbf{RMSE} & \textbf{PwRMSE} & \textbf{R$^2$} & \textbf{RMSE} & \textbf{PwRMSE} \\[4pt] 
	\hline \\[-8pt]
		\multicolumn{2}{l}{NO$_{2}$} &&&&&\\
	  \multicolumn{2}{c}{$(a)$} & 2010  & 0.74 & 7.4 & 9.6 & 0.68 & 8.2 & 10.6 \\
	                            & & 2011  & 0.75 & 7.4 & 9.6 & 0.71 & 7.9 & 9.9 \\
	                            & & 2012  & 0.76 & 6.9 & 8.8 & 0.72 & 7.4 & 9.1  \\
	                            & & 2013  & 0.78 & 6.4 & 8.1 & 0.74 & 6.8 & 8.5 \\
	                            & & 2014  & 0.78 & 6.1 & 7.9 & 0.73 & 6.6 & 8.5 \\
	                            & & 2015  & 0.77 & 6.4 & 8.3 & 0.71 & 7.2 & 8.8 \\
	                            & & 2016  & 0.77 & 6.1 & 7.9 & 0.73 & 6.6 & 8.4 \\[4pt]
	  \multicolumn{2}{c}{$(b)$} & 2010  & 0.74 & 7.9 & 10.1 & 0.68 & 8.4 & 10.8 \\
	                            & & 2011  & 0.75 & 8.0 & 10.3 & 0.71 & 8.5 & 10.8 \\
	                            & & 2012  & 0.76 & 7.0 & 8.9 & 0.72 & 7.4 & 9.2 \\
	                            & & 2013  & 0.78 & 6.4 & 8.2 & 0.74 & 6.8 & 8.6 \\
	                            & & 2014  & 0.77 & 6.4 & 8.1 & 0.73 & 6.9 & 8.6 \\
	                            & & 2015  & 0.77 & 6.5 & 8.3 & 0.71 & 7.2 & 8.7 \\
	                            & & 2016  & 0.76 & 6.3 & 8.1 & 0.73 & 6.8 & 8.7 \\[4pt]
	  \multicolumn{2}{c}{$(c)$} & 2010  & 0.49 & 10.6 & 12.5 & 0.50 & 10.4 & 12.6 \\
	                            & & 2011  & 0.48 & 11.4 & 13.2 & 0.47 & 11.3 & 13.4 \\
	                            & & 2012  & 0.50 & 10.1 & 11.7 & 0.48 & 10.1 & 11.6 \\
	                            & & 2013  & 0.50 & 9.6 & 10.9 & 0.49 & 9.7 & 10.7 \\
	                            & & 2014  & 0.52 & 9.1 & 10.2 & 0.53 & 9.0 & 10.1 \\
	                            & & 2015  & 0.51 & 9.4 & 10.8 & 0.51 & 9.5 & 10.5 \\
	                            & & 2016  & 0.52 & 8.9 & 10.2 & 0.53 & 8.8 & 10.1 \\[4pt]
		\multicolumn{2}{l}{PM$_{2.5}$} &&&&&\\
	  \multicolumn{2}{c}{$(a)$} & 2010  & 0.91 & 1.6 & 1.6 & 0.81 & 2.2 & 2.1 \\
	                            & & 2011  & 0.93 & 1.6 & 1.8 & 0.87 & 2.1 & 2.3 \\
	                            & & 2012  & 0.91 & 1.5 & 1.7 & 0.86 & 2.0 & 2.2 \\
	                            & & 2013  & 0.93 & 1.3 & 1.4 & 0.89 & 1.6 & 1.6 \\
	                            & & 2014  & 0.92 & 1.2 & 1.4 & 0.85 & 1.6 & 1.8 \\
	                            & & 2015  & 0.94 & 1.2 & 1.4 & 0.90 & 1.6 & 1.8 \\
	                            & & 2016  & 0.92 & 1.2 & 1.4 & 0.86 & 1.6 & 1.7 \\[4pt]
	  \multicolumn{2}{c}{$(b)$} & 2010  & 0.82 & 2.9 & 2.7 & 0.76 & 3.2 & 3.0 \\
	                            & & 2011  & 0.89 & 3.5 & 3.6 & 0.86 & 3.8 & 4.0 \\
	                            & & 2012  & 0.86 & 1.9 & 2.0 & 0.82 & 2.2 & 2.4 \\
	                            & & 2013  & 0.87 & 2.0 & 1.9 & 0.84 & 2.2 & 2.1 \\
	                            & & 2014  & 0.84 & 1.8 & 2.1 & 0.80 & 2.1 & 2.3 \\
	                            & & 2015  & 0.89 & 1.8 & 1.9 & 0.85 & 2.1 & 2.2 \\
	                            & & 2016  & 0.86 & 2.0 & 2.1 & 0.82 & 2.3 & 2.3 \\[4pt]
	  \multicolumn{2}{c}{$(c)$} & 2010  & 0.45 & 4.2 & 3.9 & 0.46 & 4.1 & 3.8 \\
	                            & & 2011  & 0.57 & 5.1 & 4.7 & 0.57 & 5.2 & 4.9 \\
	                            & & 2012  & 0.48 & 3.6 & 3.5 & 0.48 & 3.8 & 3.6 \\
	                            & & 2013  & 0.59 & 3.6 & 3.3 & 0.58 & 3.7 & 3.3 \\
	                            & & 2014  & 0.65 & 2.5 & 2.5 & 0.62 & 2.5 & 2.6 \\
	                            & & 2015  & 0.69 & 3.0 & 3.0 & 0.67 & 3.0 & 3.0 \\
	                            & & 2016  & 0.57 & 3.0 & 2.9 & 0.58 & 2.9 & 2.7 \\[4pt]
	\hline
	\end{tabular}
\end{table}

\begin{landscape}
\newpage
\small 
%%%%%%%%%%%%%%%%%%%%%%%%%%%%%%%%%%%%%%%%%%%%%%
%%%% Table of concentrations and exposures %%%
%%%%%%%%%%%%%%%%%%%%%%%%%%%%%%%%%%%%%%%%%%%%%%

\begin{longtable}[c]{lccccccccc}
	% Top layer of header (Pollutants)
	\caption{Estimated annual average concentrations and population-weighted concentrations of NO$_2$ and PM$_{2.5}$ (in $\mu$g/m$^3$) with associated 95\% prediction intervals between 2010 and 2016 by country, together with overall in Western Europe.}\label{tab:annavg}\\
	\hline \\[-8pt]
	& & \multicolumn{4}{c}{\textbf{NO$_2$}} & \multicolumn{4}{c}{\textbf{PM$_{2.5}$}}   \\[1pt] 
	\cmidrule(lr){3-6}
	\cmidrule(lr){7-10}
	% Middle layer of header (unweighted and population-weighted)
	& & \multicolumn{2}{c}{\textbf{Unweighted}}  & \multicolumn{2}{c}{\textbf{Population-weighted}}& \multicolumn{2}{c}{\textbf{Unweighted}}  & \multicolumn{2}{c}{\textbf{Population-weighted}}  \\[1pt] 
	\cmidrule(lr){3-4}
	\cmidrule(lr){5-6}
	\cmidrule(lr){7-8}
	\cmidrule(lr){9-10}
	% Bottom layer of header 
	\textbf{Country} & \textbf{Year} & \textbf{Median} & \textbf{Interval} & \textbf{Median} & \textbf{Interval}& \textbf{Median} & \textbf{Interval} & \textbf{Median} & \textbf{Interval} \\[3pt] 
	\hline\\[1pt]
	% Repeat this header on every page
	\endhead
	% Bottom of table 
	\\[-5pt]\hline 
	\endfoot

	Western Europe & 2010 &  8.9 & (1.2, 27.2) & 23.6 & (6.8, 51.6) &  9.9 & (2.6, 20.7) & 15.0 & (7.1, 27.0) \\ 
                   & 2011 &  8.2 & (1.1, 24.8) & 21.6 & (6.2, 49.6) &  9.8 & (2.8, 21.0) & 14.1 & (6.9, 27.3) \\ 
                   & 2012 &  9.3 & (1.5, 27.3) & 23.9 & (7.2, 53.5) & 10.7 & (2.7, 22.6) & 15.4 & (7.1, 31.7) \\ 
                   & 2013 &  8.5 & (1.2, 25.4) & 22.3 & (6.5, 50.4) &  8.6 & (2.2, 18.1) & 13.0 & (5.6, 24.3) \\ 
                   & 2014 &  7.9 & (1.0, 24.5) & 21.3 & (5.8, 49.3) &  8.9 & (3.2, 18.3) & 13.1 & (6.8, 22.9) \\ 
                   & 2015 &  7.0 & (0.8, 22.2) & 19.4 & (5.1, 45.6) &  9.0 & (2.4, 18.5) & 12.5 & (5.9, 26.0) \\ 
                   & 2016 &  7.4 & (0.9, 23.3) & 20.4 & (5.5, 47.2) &  7.9 & (2.2, 16.4) & 11.3 & (5.3, 23.1) \\[5pt] 
	Andorra        & 2010 & 12.6 & (5.3, 22.3) & 18.0 & (14.1, 22.4) &  6.0 & (4.4, 8.4) &  6.7 & (5.1, 8.9) \\ 
                   & 2011 & 12.1 & (4.9, 21.4) & 17.3 & (13.5, 21.5) &  6.3 & (4.6, 8.7) &  7.0 & (5.2, 9.3) \\ 
                   & 2012 & 13.4 & (5.7, 23.0) & 18.7 & (15.0, 23.0) &  6.9 & (4.9, 9.5) &  7.7 & (5.7, 10.2) \\ 
                   & 2013 & 13.3 & (5.5, 22.8) & 18.7 & (15.2, 23.0) &  4.7 & (3.3, 6.3) &  5.4 & (3.8, 6.8) \\ 
                   & 2014 & 11.7 & (4.6, 20.8) & 17.0 & (13.5, 21.6) &  5.2 & (3.6, 7.1) &  5.9 & (4.2, 7.5) \\ 
                   & 2015 & 11.3 & (4.4, 20.3) & 16.5 & (13.0, 20.6) &  5.4 & (3.8, 7.5) &  6.1 & (4.4, 7.9) \\ 
                   & 2016 & 11.3 & (4.4, 20.2) & 16.5 & (13.4, 20.2) &  4.2 & (2.9, 5.8) &  4.8 & (3.4, 6.2) \\[5pt] 
	Austria        & 2010 & 11.8 & (3.7, 28.4) & 22.8 & (8.5, 42.3) & 13.1 & (7.7, 20.5) & 17.5 & (11.0, 22.7) \\ 
                   & 2011 & 10.2 & (3.0, 24.8) & 20.4 & (7.3, 38.2) & 12.0 & (7.3, 19.2) & 15.9 & (10.2, 21.1) \\ 
                   & 2012 & 11.8 & (3.8, 27.7) & 22.7 & (8.6, 42.2) & 12.8 & (7.8, 20.4) & 17.1 & (10.9, 22.5) \\ 
                   & 2013 & 10.9 & (3.4, 25.8) & 21.4 & (8.0, 39.7) & 10.7 & (6.3, 17.0) & 14.7 & (9.1, 19.2) \\ 
                   & 2014 & 10.2 & (3.1, 25.0) & 20.4 & (7.3, 38.9) & 10.3 & (6.0, 16.8) & 14.2 & (8.6, 18.6) \\ 
                   & 2015 &  9.3 & (2.6, 24.2) & 19.1 & (6.5, 38.4) & 10.5 & (6.4, 16.7) & 14.0 & (9.0, 18.5) \\ 
                   & 2016 &  9.6 & (2.7, 24.2) & 19.7 & (6.8, 38.3) &  9.0 & (5.5, 14.2) & 12.2 & (7.8, 16.0) \\[5pt] 
	Belgium        & 2010 & 17.2 & (6.9, 33.3) & 25.9 & (11.6, 43.5) & 15.4 & (9.7, 22.1) & 18.1 & (11.8, 23.3) \\ 
                   & 2011 & 15.4 & (5.8, 30.7) & 23.7 & (10.1, 40.7) & 13.8 & (8.7, 19.8) & 16.3 & (10.6, 20.9) \\ 
                   & 2012 & 17.3 & (6.9, 33.9) & 26.3 & (11.5, 44.6) & 15.8 & (9.4, 22.8) & 18.6 & (11.6, 24.1) \\ 
                   & 2013 & 16.4 & (6.5, 32.3) & 25.0 & (11.0, 42.8) & 13.7 & (8.9, 19.0) & 15.9 & (10.7, 20.1) \\ 
                   & 2014 & 14.5 & (5.4, 29.5) & 22.6 & (9.5, 39.3) & 12.2 & (7.7, 17.6) & 14.6 & (9.2, 18.5) \\ 
                   & 2015 & 13.0 & (4.6, 27.3) & 20.7 & (8.4, 37.0) & 11.7 & (7.5, 16.2) & 13.7 & (9.1, 17.1) \\ 
                   & 2016 & 14.2 & (5.1, 28.9) & 22.2 & (9.2, 38.5) & 11.2 & (6.7, 15.5) & 13.0 & (8.4, 16.5) \\[5pt] 
	Denmark        & 2010 & 10.7 & (4.7, 20.9) & 16.8 & (6.6, 37.7) & 12.2 & (9.1, 15.8) & 13.0 & (9.6, 16.2) \\ 
                   & 2011 & 10.1 & (4.4, 20.0) & 16.0 & (6.2, 35.3) & 12.4 & (9.3, 16.0) & 13.5 & (10.2, 16.9) \\ 
                   & 2012 & 11.3 & (5.2, 21.7) & 17.5 & (7.1, 38.2) & 11.2 & (8.4, 14.7) & 12.3 & (9.3, 16.1) \\ 
                   & 2013 & 10.4 & (4.6, 20.6) & 16.5 & (6.5, 36.8) &  9.2 & (7.0, 12.1) & 10.1 & (7.7, 13.0) \\ 
                   & 2014 &  9.7 & (4.2, 19.3) & 15.4 & (6.0, 33.2) & 11.9 & (8.8, 15.7) & 13.0 & (9.7, 16.7) \\ 
                   & 2015 &  7.9 & (3.1, 16.8) & 13.0 & (4.6, 29.3) &  9.9 & (7.4, 12.9) & 10.8 & (8.2, 13.8) \\ 
                   & 2016 &  8.8 & (3.6, 18.1) & 14.3 & (5.3, 32.3) &  8.8 & (6.5, 11.7) &  9.7 & (7.2, 12.5) \\[5pt] 
	Finland        & 2010 &  6.7 & (1.7, 14.9) & 14.1 & (4.8, 31.3) &  6.1 & (3.1, 8.9) &  7.9 & (5.1, 10.3) \\ 
                   & 2011 &  6.6 & (1.7, 14.4) & 13.5 & (4.7, 28.5) &  5.8 & (3.3, 8.5) &  7.1 & (4.9, 8.9) \\ 
                   & 2012 &  7.4 & (2.1, 15.7) & 14.9 & (5.5, 30.8) &  5.8 & (3.1, 8.4) &  7.3 & (4.9, 9.3) \\ 
                   & 2013 &  6.4 & (1.6, 14.3) & 13.4 & (4.6, 29.2) &  4.5 & (2.5, 6.7) &  5.8 & (3.8, 7.8) \\ 
                   & 2014 &  6.1 & (1.5, 13.8) & 13.1 & (4.3, 27.9) &  6.1 & (3.4, 8.8) &  7.7 & (5.2, 9.9) \\ 
                   & 2015 &  5.5 & (1.1, 12.8) & 12.0 & (3.8, 26.2) &  4.7 & (2.7, 6.8) &  5.9 & (4.0, 7.5) \\ 
                   & 2016 &  5.8 & (1.3, 13.2) & 12.6 & (4.0, 26.8) &  4.3 & (2.5, 6.5) &  5.3 & (3.7, 7.0) \\[5pt] 
	France         & 2010 &  9.6 & (2.2, 22.3) & 19.1 & (6.2, 49.8) & 13.4 & (7.7, 18.5) & 16.3 & (9.9, 22.9) \\ 
                   & 2011 &  8.5 & (1.9, 20.3) & 17.4 & (5.4, 49.8) & 12.3 & (7.5, 17.4) & 15.1 & (9.1, 21.2) \\ 
                   & 2012 &  9.8 & (2.5, 22.3) & 19.2 & (6.5, 54.1) & 13.8 & (8.2, 19.7) & 17.2 & (10.1, 24.8) \\ 
                   & 2013 &  9.3 & (2.3, 21.6) & 18.5 & (6.1, 52.6) & 11.1 & (6.4, 16.0) & 14.0 & (8.5, 21.1) \\ 
                   & 2014 &  8.3 & (1.8, 20.2) & 17.1 & (5.2, 50.3) &  9.9 & (6.3, 14.0) & 12.3 & (7.9, 16.7) \\ 
                   & 2015 &  7.3 & (1.5, 18.5) & 15.6 & (4.5, 45.8) & 10.7 & (6.7, 14.5) & 13.0 & (8.2, 17.6) \\ 
                   & 2016 &  7.9 & (1.6, 19.5) & 16.6 & (4.9, 49.2) &  9.2 & (5.7, 13.1) & 11.6 & (7.3, 16.2) \\[5pt] 
	Germany        & 2010 & 15.4 & (7.1, 32.4) & 24.7 & (10.7, 51.1) & 14.5 & (10.5, 19.7) & 16.4 & (11.9, 22.2) \\ 
                   & 2011 & 13.7 & (6.2, 29.4) & 22.6 & (9.5, 47.4) & 13.1 & (9.4, 17.6) & 14.6 & (10.6, 19.8) \\ 
                   & 2012 & 15.5 & (7.4, 32.5) & 24.9 & (10.9, 51.9) & 13.7 & (10.0, 18.2) & 15.5 & (11.4, 20.7) \\ 
                   & 2013 & 14.7 & (6.7, 31.1) & 23.7 & (10.2, 50.3) & 11.9 & (8.9, 15.6) & 13.7 & (10.1, 17.8) \\ 
                   & 2014 & 13.6 & (6.1, 29.5) & 22.5 & (9.3, 48.8) & 13.1 & (8.8, 17.9) & 14.4 & (10.1, 20.3) \\ 
                   & 2015 & 12.2 & (5.1, 27.3) & 20.6 & (8.2, 45.5) & 11.6 & (8.6, 14.8) & 13.0 & (9.8, 16.7) \\ 
                   & 2016 & 12.8 & (5.6, 28.2) & 21.4 & (8.7, 47.0) & 10.6 & (7.7, 13.9) & 11.9 & (8.7, 16.0) \\[5pt] 
	Greece         & 2010 &  7.4 & (1.0, 20.1) & 19.7 & (4.4, 49.4) & 14.6 & (7.3, 26.7) & 19.6 & (9.8, 31.8) \\ 
                   & 2011 &  6.6 & (0.8, 18.4) & 18.1 & (3.8, 43.7) & 16.5 & (7.9, 29.6) & 21.5 & (10.6, 34.8) \\ 
                   & 2012 &  7.5 & (1.1, 19.8) & 19.4 & (4.5, 46.4) & 16.9 & (8.3, 30.8) & 22.8 & (11.3, 37.2) \\ 
                   & 2013 &  6.9 & (0.9, 18.7) & 18.2 & (4.0, 43.7) & 13.1 & (6.4, 24.0) & 17.4 & (8.7, 28.4) \\ 
                   & 2014 &  6.5 & (0.7, 18.3) & 18.2 & (3.8, 48.1) & 13.3 & (6.7, 24.6) & 17.7 & (9.1, 28.6) \\ 
                   & 2015 &  5.7 & (0.5, 16.9) & 16.6 & (3.2, 43.9) & 13.6 & (6.7, 25.4) & 18.1 & (9.1, 29.7) \\ 
                   & 2016 &  6.1 & (0.6, 17.6) & 17.6 & (3.5, 47.2) & 12.8 & (6.4, 23.8) & 17.3 & (8.7, 28.1) \\[5pt] 
	Hungary        & 2010 & 13.5 & (5.9, 23.8) & 20.3 & (8.4, 42.1) & 18.6 & (12.1, 26.5) & 21.2 & (13.5, 29.0) \\ 
                   & 2011 & 12.2 & (5.2, 21.8) & 18.6 & (7.5, 38.7) & 18.5 & (12.3, 26.4) & 21.1 & (13.7, 29.0) \\ 
                   & 2012 & 13.8 & (6.2, 23.9) & 20.5 & (8.7, 41.4) & 19.1 & (12.9, 27.3) & 21.8 & (14.3, 30.2) \\ 
                   & 2013 & 12.6 & (5.4, 22.2) & 19.0 & (7.8, 38.6) & 15.6 & (10.5, 22.5) & 17.8 & (11.7, 25.0) \\ 
                   & 2014 & 12.0 & (5.0, 21.7) & 18.4 & (7.3, 39.3) & 15.3 & (10.3, 21.7) & 17.3 & (11.5, 23.6) \\ 
                   & 2015 & 10.4 & (4.1, 19.3) & 16.2 & (6.2, 34.5) & 14.6 & (9.6, 21.3) & 16.4 & (10.8, 23.4) \\ 
                   & 2016 & 11.4 & (4.6, 20.9) & 17.6 & (6.9, 38.1) & 12.5 & (8.2, 18.4) & 14.0 & (9.2, 19.8) \\[5pt] 
	Ireland        & 2010 &  6.1 & (2.0, 13.6) & 11.1 & (3.2, 25.3) &  8.7 & (5.8, 11.7) &  9.6 & (6.9, 12.6) \\ 
                   & 2011 &  5.3 & (1.5, 12.2) &  9.7 & (2.6, 23.0) &  8.9 & (6.1, 11.9) &  9.5 & (7.1, 12.8) \\ 
                   & 2012 &  6.4 & (2.1, 13.8) & 11.2 & (3.4, 25.2) &  9.6 & (6.6, 12.8) & 10.4 & (7.7, 13.8) \\ 
                   & 2013 &  6.1 & (2.0, 13.3) & 10.7 & (3.2, 23.5) &  9.0 & (6.1, 11.7) &  9.7 & (7.3, 12.7) \\ 
                   & 2014 &  4.9 & (1.3, 11.5) &  9.1 & (2.3, 20.3) &  8.7 & (5.9, 11.4) &  9.4 & (6.9, 12.3) \\ 
                   & 2015 &  4.3 & (1.0, 10.6) &  8.3 & (1.9, 19.6) &  7.7 & (5.2, 10.1) &  8.3 & (6.1, 10.9) \\ 
                   & 2016 &  4.8 & (1.3, 11.5) &  9.1 & (2.3, 21.3) &  7.6 & (5.1, 9.9) &  8.2 & (6.1, 10.7) \\[5pt] 
	Italy          & 2010 & 13.1 & (2.5, 35.3) & 28.6 & (7.7, 59.1) & 12.1 & (6.2, 27.2) & 17.8 & (8.0, 31.7) \\ 
                   & 2011 & 11.7 & (2.1, 32.8) & 26.3 & (6.9, 58.1) & 12.6 & (6.6, 27.4) & 18.5 & (8.4, 32.1) \\ 
                   & 2012 & 13.3 & (2.7, 35.6) & 28.8 & (8.0, 61.9) & 13.7 & (7.1, 33.1) & 20.8 & (9.1, 38.7) \\ 
                   & 2013 & 11.9 & (2.2, 32.2) & 26.0 & (7.0, 57.1) & 10.9 & (5.7, 24.9) & 16.5 & (7.5, 29.5) \\ 
                   & 2014 & 11.8 & (2.0, 32.5) & 26.0 & (6.7, 57.1) & 11.1 & (6.1, 23.0) & 16.2 & (8.0, 26.5) \\ 
                   & 2015 & 10.1 & (1.5, 29.0) & 23.2 & (5.7, 52.2) & 12.5 & (6.5, 27.5) & 18.4 & (8.5, 31.4) \\ 
                   & 2016 & 10.6 & (1.6, 30.2) & 24.2 & (6.0, 53.9) & 10.7 & (5.7, 24.1) & 15.8 & (7.6, 27.9) \\ [5pt]
	Liechtenstein  & 2010 & 17.7 & (6.9, 35.2) & 27.0 & (16.7, 40.0) & 10.8 & (7.6, 15.7) & 13.6 & (10.6, 16.6) \\ 
                   & 2011 & 15.0 & (5.9, 30.6) & 23.7 & (14.5, 35.4) & 10.1 & (7.2, 14.6) & 12.7 & (10.0, 15.7) \\ 
                   & 2012 & 18.0 & (7.5, 34.9) & 26.8 & (17.2, 40.2) & 11.1 & (7.9, 16.1) & 13.9 & (11.0, 17.3) \\ 
                   & 2013 & 17.2 & (6.9, 33.6) & 26.1 & (16.4, 38.7) &  9.0 & (6.5, 12.9) & 11.4 & (8.9, 13.8) \\ 
                   & 2014 & 16.3 & (6.2, 33.3) & 25.0 & (15.4, 38.5) &  8.7 & (6.3, 12.5) & 11.0 & (8.7, 13.2) \\ 
                   & 2015 & 15.9 & (5.9, 32.8) & 24.1 & (14.8, 37.9) &  9.7 & (7.0, 13.8) & 12.2 & (9.7, 14.8) \\ 
                   & 2016 & 15.4 & (5.8, 31.5) & 24.0 & (14.6, 36.7) &  8.3 & (6.0, 11.9) & 10.6 & (8.4, 12.7) \\[5pt]
	Lithuania      & 2010 &  9.2 & (3.9, 17.1) & 15.2 & (5.7, 27.4) & 12.5 & (8.3, 17.7) & 14.5 & (9.3, 20.1) \\ 
                   & 2011 &  8.6 & (3.6, 16.0) & 14.0 & (5.3, 25.0) & 12.3 & (8.2, 17.3) & 14.0 & (9.1, 18.8) \\ 
                   & 2012 &  9.6 & (4.2, 17.4) & 15.5 & (6.1, 26.9) & 14.8 & (9.8, 21.1) & 17.0 & (10.8, 23.3) \\ 
                   & 2013 &  9.0 & (3.8, 16.5) & 14.7 & (5.6, 26.2) & 13.8 & (9.2, 19.9) & 15.8 & (10.1, 22.7) \\ 
                   & 2014 &  8.5 & (3.5, 16.0) & 14.3 & (5.2, 26.7) & 15.5 & (10.3, 22.5) & 17.5 & (11.3, 25.9) \\ 
                   & 2015 &  7.6 & (2.9, 14.6) & 12.9 & (4.5, 24.1) & 12.9 & (8.5, 19.4) & 14.6 & (9.3, 23.4) \\ 
                   & 2016 &  7.9 & (3.2, 15.2) & 13.5 & (4.8, 25.6) & 10.8 & (7.2, 16.3) & 12.2 & (7.9, 19.7) \\[5pt]
	Luxembourg     & 2010 & 14.3 & (7.8, 29.0) & 18.2 & (11.1, 26.3) & 13.2 & (10.6, 16.2) & 12.7 & (11.2, 16.8) \\ 
                   & 2011 & 13.4 & (6.8, 28.6) & 17.4 & (10.4, 24.8) & 11.8 & (9.5, 14.4) & 11.5 & (10.1, 15.0) \\ 
                   & 2012 & 14.2 & (7.7, 29.9) & 18.6 & (11.3, 26.1) & 13.1 & (10.2, 16.2) & 13.2 & (11.6, 16.7) \\ 
                   & 2013 & 14.2 & (7.6, 30.3) & 18.7 & (10.9, 26.2) & 12.2 & (9.7, 14.8) & 12.1 & (10.7, 15.1) \\ 
                   & 2014 & 12.9 & (6.5, 29.8) & 18.1 & (10.0, 24.7) & 11.0 & (8.5, 13.3) & 10.9 & (9.6, 13.8) \\ 
                   & 2015 & 12.0 & (5.8, 27.8) & 16.5 & (8.9, 23.2) & 11.0 & (8.3, 13.3) & 10.9 & (9.6, 13.8) \\ 
                   & 2016 & 12.3 & (6.1, 28.9) & 17.4 & (9.5, 24.1) &  9.9 & (7.4, 12.1) & 10.0 & (8.6, 12.4) \\[5pt]
	Monaco         & 2010 & 20.6 & (9.5, 40.0) & 27.7 & (17.4, 42.3) & 15.5 & (11.5, 21.4) & 18.1 & (13.9, 22.6) \\ 
                   & 2011 & 20.9 & (10.2, 41.4) & 28.6 & (17.7, 44.3) & 15.9 & (12.1, 22.2) & 18.7 & (14.4, 23.4) \\ 
                   & 2012 & 24.8 & (12.7, 46.3) & 32.9 & (21.1, 50.8) & 18.2 & (13.6, 25.3) & 21.3 & (16.2, 26.6) \\ 
                   & 2013 & 21.7 & (10.5, 42.3) & 29.4 & (18.2, 46.1) & 13.0 & (9.9, 18.1) & 15.3 & (11.9, 19.2) \\ 
                   & 2014 & 22.0 & (10.5, 42.9) & 29.5 & (18.4, 46.8) & 11.9 & (9.2, 16.7) & 14.1 & (11.0, 17.5) \\ 
                   & 2015 & 18.9 & (8.5, 38.4) & 26.1 & (15.7, 42.3) & 13.1 & (9.9, 18.4) & 15.5 & (11.7, 19.6) \\ 
                   & 2016 & 20.2 & (9.2, 40.3) & 27.4 & (16.8, 43.9) & 12.1 & (9.0, 17.0) & 14.3 & (10.8, 18.6) \\[5pt]
	Netherlands    & 2010 & 17.7 & (9.6, 35.4) & 27.3 & (12.8, 44.4) & 15.7 & (12.4, 20.1) & 16.9 & (13.4, 20.9) \\ 
                   & 2011 & 16.6 & (9.2, 33.2) & 25.6 & (12.3, 42.3) & 14.1 & (11.3, 18.1) & 15.2 & (12.2, 18.8) \\ 
                   & 2012 & 18.4 & (10.4, 36.0) & 27.9 & (13.6, 45.4) & 14.9 & (11.6, 19.8) & 16.2 & (12.6, 20.2) \\ 
                   & 2013 & 16.6 & (9.1, 33.5) & 25.7 & (12.2, 42.2) & 12.7 & (9.8, 16.6) & 13.9 & (10.8, 17.0) \\ 
                   & 2014 & 15.5 & (8.5, 31.7) & 24.3 & (11.4, 40.6) & 13.4 & (10.9, 16.9) & 14.2 & (11.5, 17.7) \\ 
                   & 2015 & 13.9 & (7.3, 29.1) & 22.2 & (10.0, 37.3) & 11.6 & (9.3, 14.8) & 12.5 & (10.0, 15.4) \\ 
                   & 2016 & 14.8 & (7.9, 30.5) & 23.4 & (10.7, 39.0) & 10.3 & (8.2, 13.4) & 11.0 & (8.9, 13.7) \\[5pt]
	Norway         & 2010 &  4.7 & (0.2, 19.5) & 21.3 & (4.1, 46.6) &  5.1 & (2.3, 10.0) &  9.6 & (4.5, 13.9) \\ 
                   & 2011 &  4.7 & (0.2, 18.7) & 20.2 & (4.0, 43.7) &  5.4 & (2.5, 10.0) &  9.0 & (4.9, 13.2) \\ 
                   & 2012 &  5.4 & (0.3, 20.3) & 22.0 & (4.7, 47.6) &  5.1 & (2.4, 9.7) &  8.9 & (4.7, 13.2) \\ 
                   & 2013 &  4.7 & (0.2, 18.6) & 20.1 & (4.1, 43.0) &  4.1 & (2.0, 8.0) &  7.6 & (3.9, 11.2) \\ 
                   & 2014 &  4.5 & (0.2, 18.4) & 20.0 & (3.9, 43.1) &  5.3 & (2.8, 9.8) &  9.1 & (5.0, 13.0) \\ 
                   & 2015 &  3.9 & (0.1, 17.2) & 18.7 & (3.3, 40.2) &  4.1 & (2.1, 8.1) &  7.6 & (3.9, 10.9) \\ 
                   & 2016 &  4.3 & (0.1, 18.0) & 19.6 & (3.6, 42.3) &  3.6 & (1.9, 7.2) &  6.8 & (3.3, 9.8) \\[5pt]
	Portugal       & 2010 &  6.5 & (2.0, 20.8) & 17.6 & (4.5, 40.2) &  7.6 & (4.7, 12.0) &  8.7 & (5.4, 14.1) \\ 
                   & 2011 &  6.0 & (1.7, 19.5) & 16.6 & (4.0, 37.3) &  7.4 & (4.7, 11.3) &  8.5 & (5.4, 13.3) \\ 
                   & 2012 &  6.8 & (2.2, 20.9) & 18.0 & (4.8, 41.0) &  8.1 & (5.1, 12.6) &  9.4 & (5.9, 15.0) \\ 
                   & 2013 &  6.0 & (1.8, 18.4) & 15.7 & (4.1, 35.7) &  6.4 & (4.0, 9.3) &  7.3 & (4.6, 11.5) \\ 
                   & 2014 &  5.5 & (1.5, 17.5) & 14.9 & (3.6, 34.7) &  7.0 & (4.1, 10.1) &  7.7 & (4.6, 11.6) \\ 
                   & 2015 &  5.0 & (1.2, 17.0) & 14.2 & (3.2, 34.2) &  8.1 & (4.7, 12.3) &  9.1 & (5.1, 14.7) \\ 
                   & 2016 &  5.0 & (1.3, 16.8) & 14.2 & (3.3, 33.5) &  6.9 & (4.0, 10.2) &  7.7 & (4.5, 12.2) \\[5pt]
	San Marino     & 2010 &  9.8 & (4.6, 15.8) & 13.8 & (6.7, 19.0) & 14.2 & (12.0, 17.0) & 15.9 & (13.0, 18.2) \\ 
                   & 2011 &  7.7 & (3.5, 13.4) & 12.0 & (5.2, 16.0) & 14.6 & (12.3, 17.3) & 16.4 & (13.5, 18.7) \\ 
                   & 2012 &  8.6 & (4.1, 14.7) & 13.2 & (6.0, 17.4) & 16.1 & (13.5, 19.3) & 18.3 & (14.8, 20.9) \\ 
                   & 2013 &  8.0 & (3.6, 13.4) & 12.0 & (5.5, 16.1) & 12.1 & (10.3, 14.4) & 13.8 & (11.2, 15.7) \\ 
                   & 2014 &  7.9 & (3.5, 13.2) & 11.7 & (5.2, 16.0) & 12.2 & (10.3, 14.5) & 13.7 & (11.3, 15.5) \\ 
                   & 2015 &  6.6 & (2.8, 11.5) & 10.2 & (4.3, 13.9) & 14.3 & (12.0, 17.1) & 16.2 & (13.2, 18.5) \\ 
                   & 2016 &  7.1 & (3.1, 12.2) & 10.9 & (4.6, 14.9) & 12.0 & (10.1, 14.4) & 13.7 & (11.1, 15.5) \\[5pt]
	Spain          & 2010 &  7.5 & (1.9, 20.1) & 23.4 & (5.9, 51.2) &  9.0 & (5.5, 15.0) & 12.1 & (7.6, 19.5) \\ 
                   & 2011 &  6.5 & (1.5, 17.9) & 21.4 & (5.2, 49.0) &  9.0 & (5.6, 15.0) & 11.9 & (7.6, 19.1) \\ 
                   & 2012 &  7.7 & (2.0, 19.9) & 23.6 & (6.1, 52.3) & 10.2 & (6.1, 17.3) & 14.1 & (8.5, 23.4) \\ 
                   & 2013 &  6.8 & (1.7, 18.3) & 21.6 & (5.4, 47.6) &  7.5 & (4.6, 13.3) & 10.0 & (6.3, 17.4) \\ 
                   & 2014 &  6.2 & (1.3, 17.2) & 20.7 & (4.8, 45.8) &  8.5 & (5.2, 14.9) & 11.1 & (7.2, 18.9) \\ 
                   & 2015 &  5.7 & (1.1, 16.6) & 19.8 & (4.3, 45.8) &  9.1 & (5.5, 15.9) & 11.8 & (7.6, 20.5) \\ 
                   & 2016 &  5.9 & (1.2, 16.7) & 20.2 & (4.5, 45.9) &  7.6 & (4.5, 13.6) & 10.3 & (6.4, 17.3) \\[5pt]
	Sweden         & 2010 &  6.1 & (0.6, 16.4) & 16.1 & (5.0, 36.5) &  4.2 & (2.0, 9.3) &  7.5 & (3.5, 11.7) \\ 
                   & 2011 &  6.2 & (0.6, 16.2) & 15.9 & (5.0, 34.7) &  4.4 & (2.1, 9.8) &  7.4 & (3.6, 12.6) \\ 
                   & 2012 &  6.9 & (0.9, 17.5) & 17.2 & (5.7, 37.1) &  4.2 & (2.0, 9.2) &  7.0 & (3.4, 11.9) \\ 
                   & 2013 &  6.1 & (0.6, 16.2) & 15.9 & (5.0, 35.6) &  3.3 & (1.7, 7.3) &  5.5 & (2.8, 9.4) \\ 
                   & 2014 &  5.8 & (0.5, 15.6) & 15.2 & (4.6, 34.0) &  4.8 & (2.5, 9.7) &  7.5 & (4.2, 12.2) \\ 
                   & 2015 &  5.0 & (0.3, 14.2) & 13.8 & (3.9, 31.9) &  3.6 & (1.8, 7.8) &  5.9 & (3.1, 10.1) \\ 
                   & 2016 &  5.4 & (0.4, 15.0) & 14.5 & (4.3, 32.3) &  3.2 & (1.7, 7.0) &  5.5 & (2.8, 9.1) \\[5pt]
	Switzerland    & 2010 &  9.5 & (1.3, 25.8) & 21.2 & (8.3, 38.3) & 10.9 & (6.4, 16.8) & 14.2 & (10.3, 19.3) \\ 
                   & 2011 &  8.5 & (1.0, 23.7) & 19.5 & (7.4, 36.1) & 10.5 & (6.6, 15.8) & 13.2 & (9.8, 18.6) \\ 
                   & 2012 & 10.1 & (1.5, 26.0) & 21.5 & (8.7, 38.7) & 11.3 & (6.7, 17.3) & 14.4 & (10.6, 20.5) \\ 
                   & 2013 &  9.7 & (1.4, 25.6) & 21.0 & (8.4, 38.4) &  9.1 & (5.1, 14.2) & 12.2 & (8.7, 16.5) \\ 
                   & 2014 &  8.4 & (1.0, 24.0) & 19.5 & (7.3, 36.6) &  8.5 & (5.0, 12.7) & 11.0 & (8.2, 14.7) \\ 
                   & 2015 &  7.6 & (0.7, 22.5) & 18.2 & (6.5, 34.7) &  9.5 & (5.5, 14.4) & 12.2 & (9.0, 16.9) \\ 
                   & 2016 &  7.9 & (0.8, 23.1) & 18.8 & (6.8, 35.4) &  8.2 & (4.7, 12.5) & 10.7 & (7.9, 14.6) \\[5pt]
	United Kingdom & 2010 & 12.5 & (3.4, 32.9) & 26.9 & (9.7, 52.8) &  9.3 & (3.9, 14.5) & 12.6 & (7.7, 16.9) \\ 
                   & 2011 & 11.3 & (3.0, 30.0) & 24.6 & (8.6, 50.5) &  9.5 & (4.5, 14.4) & 12.5 & (7.9, 16.9) \\ 
                   & 2012 & 12.8 & (3.7, 33.4) & 27.4 & (10.0, 55.4) & 10.2 & (4.7, 15.9) & 13.6 & (8.5, 19.1) \\ 
                   & 2013 & 11.9 & (3.4, 31.1) & 25.4 & (9.2, 51.8) &  9.0 & (3.8, 13.3) & 11.7 & (7.3, 15.9) \\ 
                   & 2014 & 10.8 & (2.7, 29.7) & 24.2 & (8.1, 49.4) &  9.4 & (4.3, 14.2) & 12.4 & (7.9, 16.4) \\ 
                   & 2015 &  9.6 & (2.3, 26.6) & 21.7 & (7.0, 45.7) &  7.9 & (3.5, 11.8) & 10.3 & (6.5, 13.5) \\ 
                   & 2016 & 10.3 & (2.4, 28.5) & 23.2 & (7.7, 46.1) &  7.6 & (2.9, 11.6) & 10.1 & (5.5, 13.6) \\[5pt]
\end{longtable}

\newpage 

\begin{longtable}[c]{lcccccccc}
\caption{Estimated changes in annual average NO$_2$ and PM$_{2.5}$ (in $\mu$g/m$^3$) between 2010 and 2016 by country, together with overall change in Western Europe.}\label{tab::summdiff} \\
	\hline \\[-8pt]
	  &  \multicolumn{4}{c}{\textbf{NO$_2$}} & \multicolumn{4}{c}{\textbf{PM$_{2.5}$}}   \\[1pt] 
	 \cmidrule(lr){2-5}
	 \cmidrule(lr){6-9}
	  &  \multicolumn{2}{c}{\textbf{Unweighted}}  & \multicolumn{2}{c}{\textbf{Population-weighted}}& \multicolumn{2}{c}{\textbf{Unweighted}}  & \multicolumn{2}{c}{\textbf{Population-weighted}}  \\[1pt] 
	 \cmidrule(lr){2-3}
	 \cmidrule(lr){4-5}
	 \cmidrule(lr){6-7}
	 \cmidrule(lr){8-9}
	\textbf{Country} &  \textbf{Median} & \textbf{Interval} & \textbf{Median} & \textbf{Interval}& \textbf{Median} & \textbf{Interval} & \textbf{Median} & \textbf{Interval} \\[3pt] 
  \hline\\[1pt]
  \endhead
  
\\[-5pt]\hline 
\endfoot
	Western Europe & -1.4 & (-4.6,  0.0) & -2.9 & (-7.4, -0.3) & -2.0 & (-6.5,  0.5) & -3.3 & (-7.7,  0.5) \\
	Andorra & -1.3 & (-2.7, -0.4) & -1.6 & (-2.8, -0.5) & -1.8 & (-3.1, -0.9) & -2.0 & (-3.3, -1.1) \\ 
	Austria & -2.2 & (-5.3, -0.5) & -2.8 & (-6.3, -0.7) & -3.9 & (-7.3, -1.7) & -5.2 & (-8.2, -2.4) \\  
	Belgium & -2.9 & (-5.4, -1.3) & -3.7 & (-6.3, -1.6) & -4.2 & (-7.6, -2.3) & -4.9 & (-8.0, -2.7) \\ 
	Denmark & -1.8 & (-3.5, -0.7) & -2.5 & (-6.0, -0.9) & -3.3 & (-5.9, -1.3) & -3.1 & (-5.8, -1.1) \\ 
	Finland & -0.8 & (-2.2, -0.1) & -1.5 & (-5.4,  0.0) & -1.7 & (-3.5,  0.2) & -2.6 & (-4.2, -0.4) \\ 
	France & -1.6 & (-3.7, -0.2) & -2.0 & (-4.8,  0.7) & -4.1 & (-6.8, -1.1) & -4.6 & (-8.1, -1.4) \\ 
	Germany & -2.5 & (-5.3, -0.9) & -3.1 & (-6.6, -0.6) & -3.9 & (-6.7, -1.9) & -4.5 & (-7.3, -2.2) \\ 
	Greece & -1.2 & (-3.3, -0.1) & -1.6 & (-4.3,  0.3) & -1.5 & (-7.7,  2.8) & -2.2 & (-9.4,  3.3) \\ 
	Hungary & -2.0 & (-3.9, -0.7) & -2.6 & (-5.3, -0.9) & -6.0 & (-11.1, -1.6) & -7.1 & (-12.8, -1.9) \\ 
	Ireland & -1.2 & (-2.7, -0.4) & -2.0 & (-4.7, -0.6) & -1.0 & (-3.3,  0.5) & -1.4 & (-3.5,  0.4) \\ 
	Italy & -2.4 & (-6.7, -0.5) & -3.7 & (-9.1, -1.0) & -1.6 & (-4.9,  1.2) & -2.0 & (-6.0,  1.4) \\ 
	Liechtenstein & -2.3 & (-4.3, -0.5) & -2.9 & (-4.8, -0.8) & -2.4 & (-4.6, -0.9) & -3.0 & (-5.1, -1.2) \\ 
	Lithuania & -1.3 & (-2.6, -0.4) & -1.6 & (-3.5, -0.4) & -1.6 & (-4.8,  2.1) & -1.9 & (-5.6,  2.8) \\ 
	Luxembourg & -1.7 & (-3.2,  1.4) & -1.0 & (-4.2,  1.5) & -3.4 & (-4.9, -1.8) & -2.8 & (-4.9, -1.2) \\ 
	Monaco & -0.5 & (-2.4,  1.7) & -0.1 & (-2.6,  2.1) & -3.3 & (-6.1, -0.8) & -3.8 & (-6.7, -0.8) \\ 
	Netherlands & -2.9 & (-5.6, -1.2) & -3.7 & (-6.8, -1.6) & -5.3 & (-7.8, -3.3) & -5.8 & (-8.3, -3.7) \\ 
	Norway & -0.4 & (-2.2,  0.3) & -1.6 & (-5.4,  0.4) & -1.5 & (-3.6,  0.2) & -2.7 & (-5.1, -0.5) \\ 
	Portugal & -1.4 & (-4.5, -0.5) & -3.3 & (-7.5, -0.8) & -0.7 & (-3.5,  1.0) & -1.0 & (-4.2,  1.1) \\ 
	San Marino & -2.6 & (-4.2, -1.3) & -2.9 & (-4.5, -1.7)  & -2.1 & (-3.9, -1.1) & -2.2 & (-4.1, -1.0) \\ 
	Spain & -1.5 & (-4.0, -0.3) & -2.9 & (-8.2, -0.5) & -1.3 & (-3.7,  0.8) & -1.7 & (-4.8,  1.2) \\ 
	Sweden & -0.6 & (-2.1,  0.1) & -1.6 & (-5.3,  0.1) & -0.9 & (-2.9,  0.2) & -1.9 & (-3.5, -0.3) \\ 
	Switzerland & -1.5 & (-4.0, -0.2) & -2.3 & (-5.3, -0.5) & -2.7 & (-5.3, -1.1) & -3.4 & (-6.1, -1.4) \\ 
	United Kingdom & -2.1 & (-5.3, -0.6) & -3.6 & (-8.0, -1.1)& -1.8 & (-4.1,  0.1) & -2.5 & (-5.1, -0.3) \\[5pt] 
\end{longtable}

\end{landscape}

\end{appendix}

\end{document}